\documentclass[draftcls,onecolumn]{IEEEtran}
\usepackage{cite}
\usepackage{url}
\ifCLASSINFOpdf
  \usepackage[]{hyperref}
  \usepackage[pdftex]{graphicx}
  \DeclareGraphicsExtensions{.pdf}
\else
  \usepackage[dvips]{graphicx}
  \DeclareGraphicsExtensions{.eps}

\fi

\usepackage{multirow}

\usepackage{tikz}
\usetikzlibrary{arrows,shapes,backgrounds,plotmarks,patterns,calc}

\tikzstyle{vnode} = [circle,draw=black,thick,inner
 sep=0pt,minimum size=5pt]
\tikzstyle{cnode} = [draw=black,thick,inner
 sep=0pt,minimum size=5pt]
\tikzstyle{dnode} = [diamond,draw=black, thick,inner
 sep=0pt,minimum size=5pt]

\usepackage{pgfplots}

\usepackage{amsthm}
\usepackage[cmex10]{amsmath}
\usepackage{amssymb}
\usepackage[font=footnotesize]{caption}
\usepackage[font=footnotesize]{subcaption}

\newtheorem{theorem}{Theorem}

\newtheorem{lemma}[theorem]{Lemma}
\newtheorem{proposition}[theorem]{Proposition}
\theoremstyle{definition}

\theoremstyle{remark}
\newtheorem{remark}{Remark}

\newcommand{\ff}{\mathbb{F}}
\newcommand{\va}{\text{A}}
\newcommand{\vb}{\text{B}}
\newcommand{\vc}{\text{C}}

\newcommand{\rank}{\mathrm{rk}}
\newcommand{\E}{\mathbb{E}}

\newcommand{\lspan}[1]{\langle#1\rangle}

\begin{document}
\title{Linearly-Coupled Fountain Codes}
\author{Shenghao~Yang,
  Soung~Chang~Liew,
  Lizhao~You
  and~Yi~Chen%
  \thanks{This paper will be presented in part at 2014 IEEE Information Theory Workshop.}%
  \thanks{S.~Yang is with Institute for Interdisciplinary Information Sciences, Tsinghua University,
    Beijing, China. Email: shyang@tsinghua.edu.cn}%
  \thanks{S.~C.~Liew and L.~You are with Department of Information Engineering, The Chinese University of
    Hong Kong, Hong Kong, China. Email: soung@ie.cuhk.edu.hk, yl013@ie.cuhk.edu.hk}%
  \thanks{Y.~Chen is with the Department of Electronic Engineering, City University of Hong
    Kong, Hong Kong, China: Email: chelseachenyi@gmail.com}%
}

\maketitle

\begin{abstract}
  Network-coded multiple access (NCMA) is a communication scheme for
  wireless multiple-access networks where physical-layer network
  coding (PNC) is employed. In NCMA, a user encodes and spreads its
  message into multiple packets. Time is slotted and multiple users
  transmit packets (one packet each) simultaneously in each
  timeslot. A sink node aims to decode the messages of all the users
  from the sequence of receptions over successive timeslots.  For
  each timeslot, the NCMA receiver recovers multiple linear
  combinations of the packets transmitted in that timeslot, forming a
  system of linear equations. Different systems of linear equations
  are recovered in different timeslots.  A message decoder then
  recovers the original messages of all the users by jointly solving
  multiple systems of linear equations obtained over different
  timeslots. We propose a low-complexity digital fountain approach for
  this coding problem, where each source node encodes its message into
  a sequence of packets using a fountain code. The aforementioned
  systems of linear equations recovered by the NCMA receiver
  effectively couple these fountain codes together.  We refer to the
  coupling of the fountain codes as a linearly-coupled (LC)
  fountain code. The ordinary belief propagation (BP) decoding
  algorithm for conventional fountain codes is not optimal for LC
  fountain codes. We propose a batched BP decoding algorithm and
  analyze the convergence of the algorithm for general LC fountain
  codes.  We demonstrate how to optimize the degree distributions and
  show by numerical results that the achievable rate region is nearly
  optimal. Our approach significantly reduces the decoding complexity
  compared with the previous NCMA schemes based on Reed-Solomon codes
  and random linear codes, and hence has the potential to increase
  throughput and decrease delay in computation-limited NCMA systems.
\end{abstract}

\section{Introduction}

Consider a wireless multiple-access network where $L$ source nodes
(users) deliver information to a sink node through a common wireless
channel. Each source node encodes its
message into multiple packets and transmits these packets sequentially
over successive timeslots. All the transmissions start at the beginning
of a timeslot, and the timeslots are long enough to complete the
transmission of a packet. %

Multiple access in such scenarios, where the goal of the sink node is
to decode the messages of all source nodes, can benefit from
\emph{physical-layer network coding (PNC)} \cite{Zhang06} (also known
as \emph{compute-and-forward} \cite{Nazer11}) by decoding linear
combinations of the packets simultaneously transmitted 
in each timeslot. Such a multiple-access scheme is called
\emph{network-coded multiple access (NCMA)} and has been studied in
\cite{lu13n,ncma14,cocco2014}, where both PNC and multiuser decoders
are employed at the physical layer to obtain the aforementioned linear
combinations.  Specifically, Lu, You and Liew \cite{lu13n}
demonstrated by a prototype that a PNC decoder can successfully
recover linear combinations of the packets while the traditional
multiuser decoder \cite{verdu1998multiuser} that does not make use of
PNC fails. %

The ultimate goal of a multiple-access network is to recover the original
messages of all users, rather than just the linear combinations of the
transmitted packets among different users. Message decoding is
hence required by NCMA to recover the original messages of all
users. In this paper, we study this message coding problem induced
by NCMA, illustrated as follows by a two-user multiple-access
network.

\subsection{Network-Coded Multiple Access with Two Users}
\label{sec:netw-coded-mult}

Consider a wireless multiple-access network with two source
nodes $\va$ and $\vb$. Nodes $\va$ and $\vb$ transmit
packets $v_\va$ and $v_\vb$ simultaneously, and the sink node receives
a superposition of the waveforms transmitted by both users. In the NCMA
scheme in \cite{lu13n}, two types of physical-layer
decoders are used to decode the received waveform: 1) a conventional
multiuser decoder that attempts to decode both $v_\va$ and $v_\vb$;
and 2) a PNC decoder that attempts to decode $v_\va + v_\vb$ (the sum
is bit-wise exclusive-or), referred to as a \emph{coupled} packet.
The combined decoding outcomes can be grouped into five events: i)
only $v_\va$ is decoded; ii) only $v_\vb$ is decoded; iii) only
$v_\va + v_\vb$ is decoded; iv) both $v_\va$ and $v_\vb$ are
decoded;\footnote{If $v_\va$ and $v_\va+v_\vb$ are decoded, we
  consider $v_\va$ and $v_\vb$ as being decoded since $v_\vb =
  v_\va+(v_\va+v_\vb)$.} and v) nothing is decoded. Experiments on
the NCMA prototype~\cite{lu13n} indicated that all the five events
have non-negligible probabilities.

Suppose that each source node has a message formed by $K$ input packets.  The source node
$\va$ ($\vb$) encodes its input packets to a sequence of coded packets
$v_\va[i]$ ($v_\vb[i]$), $i=1,\ldots, N$ using an erasure-correction code,
where $N$ is the block length of the code. Source nodes $\va$ and $\vb$
transmit packets $v_\va[i]$ and $v_\vb[i]$ simultaneously to the sink node.
According to the five events above, the outputs of the physical layer
of the sink node can be put into three groups. Specifically,  for
certain subsets $I_1,I_2,I_3\subset\{1,2,\ldots,N\}$ with $(I_1\cup
I_2)\cap I_3=\emptyset$, the three groups are
\begin{equation}\label{eq:12}
  \{v_\va[i], i\in I_1\}, \{v_\vb[i], i\in I_2\} \text{ and } \{v_\va[i]+v_\vb[i], i\in I_3\},
\end{equation}
where the first group is the coded packets of source node $\va$,
the second group is the coded packets of source node $\vb$ and
the third group is the coupled packets. 

A natural question that arises is how to encode at the source nodes so that the sink node
in NCMA can decode the input packets of all the source nodes reliably
using the output packets in \eqref{eq:12}.
In \cite{ncma14}, Reed-Solomon codes and uniform random linear codes
are used to encode the input packets at the source node. The output
packets categorized by the three groups are treated
as a coupling of two Reed-Solomon codes (or two uniform random linear
codes). The two coupled
Reed-Solomon codes (uniform random linear codes) can be decoded
jointly by a unified equation system, which is optimal in the sense
that as long as there are enough linearly independent equations,
the input packets of both source nodes can be decoded \cite{ncma14}.

The joint decoding of the coupled Reed-Solomon codes (uniform random
linear codes), however, is complex. The decoding complexity by using
Gaussian elimination is of $O((2K)^3+(2K)^2T)$ finite-field
operations, where $T$ is the number of field elements in a packet. As
a result, the system prototype in \cite{ncma14} can only demonstrate
the real-time decoding for low data rates. Further, if NCMA is
generalized to accommodate more than two source nodes, the decoding
complexity will be much higher. Take an $L$-user NCMA system for example,
using Reed-Soloman codes (uniform random linear codes) may result in a
decoding complexity of $O(L^3K^3+L^2K^2T)$ finite-field operations,
making real-time decoding even more challenging. This observation
motivates us to study a more efficient coding scheme for NCMA with low encoding/decoding complexity.

\subsection{Paper Contributions}

For a general NCMA system with $L\geq 2$ users, the sink node can
decode as many as $L$ linear combinations with coefficients over a
finite field for a set of simultaneously transmitted
packets in each timeslot.\footnote{PNC can also operate over a finite ring
  \cite{feng2013algebraic}. Readers can refer to
  \cite{feng2013algebraic,feng2013communication} to see how to use
  finite rings in PNC and how to extend the results over finite fields
  to finite rings.} In this paper, we study how to efficiently recover
the original messages of all the users using the linear combinations
decoded in different timeslots. This message coding problem induced
by NCMA is the channel coding for linear multiple-access
channels (MACs), where the output is a set of linear combinations of
the multiple input packets.

Fountain codes (e.g., LT codes \cite{lubyLT} and Raptor codes
\cite{shokRaptor}) were originally introduced for erasure channels and
have the advantages of ratelessness and low encoding/decoding
complexity.  We propose a digital fountain approach for NCMA, where
each user encodes its $K$ input packets using a fountain code.  These
linear combinations decoded by the physical layer of the sink node
over a number of timeslots are collectively called a
\emph{linearly-coupled (LC) fountain code}. We use \emph{LC-$L$} to
indicate the LC fountain code involving $L$ users.

The ordinary BP decoding algorithm of fountain codes is not optimal
for LC fountain codes, except for the case of two users. We instead
propose a \emph{batched BP decoding} algorithm, which processes the
linear combinations decoded from the same timeslot
jointly (see Section~\ref{sec:gener-batch-bp}). The decoding
complexity of batched BP decoding is of $O(LK(\tilde L^2+LT))$
finite-field operations, where $\tilde L \leq L$ is the maximum number
of linearly independent
combinations that can be decoded by the physical-layer for a
single timeslot. The batched BP decoding can be regarded as
the combination of local Gaussian elimination and the ordinary BP
decoding. %
We analyze the performance of the batched BP
decoding algorithm by performing these two parts
iteratively (Theorem~\ref{the:L}).

The degree distributions of the original fountain codes designed for
the single-user erasure channel is far from optimal for the linear
multiple-access channel.  We provide a geometric analysis of the
convergence of the batched BP decoding (Theorem~\ref{the:L2}).  This
convergence analysis induces the optimization problems of the degree
distributions of the LC fountain codes.  We use binary LC-2 and LC-3
fountain codes to illustrate how to optimize the degree distributions.
Since each user has an achievable rate, we formulate two degree distribution
optimization problems. The first aims
to maximize one user's rate given that the other users' rates are
fixed. The second aims to maximize the sum rate of all users. We
solve these optimization problems numerically. Our numerical
results show that binary LC-2 and LC-3 fountain codes can achieve a
rate region close to the capacity region of the linear MAC induced by
NCMA.

\subsection{Other Related Works}

This paper assumes that the PNC decoder can reliably recover one or
more linear combinations of the packets transmitted
simultaneously. The decoding of the XOR of the packets of two users
has been extensively investigated \cite{shengli09,wubben10} (see also
the overview \cite{liew2013physical}). The decoding of multiple linear
combinations over a larger alphabet has been studied in \cite{Nazer11,
  feng2013algebraic}. Our work in this paper can be applied to NCMA
with various PNC schemes.

Zhu and Gastpar \cite{zhu2014isit,zhu2014multiple} recently studied
the achievable rate region of Gaussian multiple-access channels by
using only a modified compute-and-forward decoder to decode linear
combinations of the messages, where the channel gains are known to the
transmitters. For a multiple-access channel of $L$
users, their scheme needs to recover $L$ linearly independent
combinations of the $L$ users' messages. By contrast, in NCMA, it is not
necessary for the physical layer to decode $L$ linearly independent
combinations for each timeslot. The message coding scheme studied
in this paper can recover the original messages of all users
from the linear combinations decoded in multiple timeslots.
Puducher, Kliewer and Fuja \cite{puducheri2007design} studied
distributed LT codes for a multiple-access relay network, where the
relay node does not receive linear combinations of the packets of the
source nodes from the physical layer. They study how to selectively
combine the
packets received from different source nodes 
so that the degree distribution observed by the sink node approximates
a robust soliton distribution.  As \cite{shengli09,wubben10}, Hern and
Narayanan \cite{hern12} also studied PNC for the two-user binary
linear MAC, wherein the purpose was to decode the XOR of the packets
of the two users. By contrast, for the application of LC fountain
codes in NCMA here, we want to recover the input packets of both
users.

Another line of works with flavors similar to ours is the study of
slotted ALOHA with successive interference
cancellation~\cite{casini2007contention,liva2011graph,paolini2011high,stefanovic2012frameless,narayanan2012iterative,paolini2014coded}.
In these works, if only one user transmits at a timeslot, the packet
can be correctly received; if multiple users transmit at the same time
slot, the sink node receives a \emph{collision}, which can be regarded
as \emph{one} linear combination of all the packets transmitted. In
NCMA, however, the sink node can recover \emph{more than one}
independent linear combinations from the collision, so that the
essential coding problem is more complicated: in particular, the ordinary BP decoding
for erasure chanels is not optimal and the ordinary tree-based
analysis of BP decoding cannot be directly applied.

\section{Problem Formulation}
\label{sec:problem-formulation}

\subsection{NCMA with Fountain Codes}

Fix two positive integers $L$ and $T$. Let $\Theta$ be an \emph{ordered} set of $L$
symbols (e.g., $\va, \vb, \vc$, and so on). 
Consider an NCMA system
with $L$ source nodes (users), each of which is labelled by a symbol
in $\Theta$. Fix a finite field $\ff_q$ of $q$ elements, called the \emph{base field}
and a degree $m$ extension field $\ff_{q^m}$.  For $s\in \Theta$, source
node $s$ has $K_s$ input packets, called the $s$-input packets.  All
the packets are regarded as column vectors of $T$ symbols in $\ff_{q^m}$.
Each source node $s$ encodes its input packets using an LT code with
degree distribution $\Psi_{s}=(\Psi_{s}[i],i=1,\ldots,D)$, where $D$
is the maximum degree. To encode the $s$-input packets, the LT-code
encoder first obtains a degree $d$ by sampling the degree distribution
$\Psi_{s}$ and then combines $d$ packets chosen uniformly at random from all the
$s$-input packets into a coded packet. The generated packet is called an $s$-coded
packet. All the $s$-coded packets are generated independently.

All the source nodes transmit the coded packets simultaneously using a
common wireless channel.  Let $v_s$ be the coded packet transmitted by
the source node $s$, $s\in\Theta$, in a timeslot. The physical-layer
decoder of the sink node tries to decode multiple linear combinations
of $v_s, s\in\Theta$ with coefficients over the base field
$\ff_q$. Suppose that $B$ linearly independent combinations are
decoded ($B$ may vary from timeslot to timeslot). They
can be expressed as
\begin{equation}\label{eq:batch}
  [v_s,s\in\Theta] H = [u_1,\ldots,u_B],
\end{equation}
where $H$ is an $L\times B$ matrix over $\ff_q$, called the
\emph{transfer matrix}, and $[v_s,s\in\Theta]$ is the matrix formed by
juxtaposing the vectors $v_s$, where $v_{s'}$ comes before $v_{s''}$
whenever $s'<s''$.

Note that in \eqref{eq:batch}, the algebraic operations are over the
field $\ff_{q^m}$.  We call the set of packets $\{u_1,\ldots,u_B\}$
decoded in a timeslot a \emph{batch}. We say that the batch is
generated by $\{v_s,s\in\Theta\}$ and packet $v_s$ is the $s$-coded
packets embedded in the batch. We assume that each coded packet is
only transmitted once. In other words, each coded packet is only
embedded in one batch.  Different batches may have different generator
matrices.

The packets decoded by the physical layer of the sink node from $N$
timeslots are collectively called an \emph{Linearly-Coupled
  (LC) fountain code formed by the coupling of $L$ fountain codes}, or
an \emph{LC-$L$ fountain code}, where $N$ is called the block-length
of the code. We assume that the empirical distribution of the transfer
matrices converges to $g$, i.e., denoting the transfer matrix of the
$i$-th batch as $H^{(i)}$,
\begin{equation*}
  \frac{|\{i:1\leq i\leq N, H^{(i)}=H\}|}{N} \rightarrow g(H),\quad \text{as $N$ tends to infinity},
\end{equation*}
where the domain of $g$ is the collection of all the full-column-rank,
$L$-row matrices over $\ff_q$ (note: this includes all such matrices
with $B$ columns, $B = 1, \ldots, L$, and an empty matrix when nothing
is decoded).

Fix $0< \eta_s < 1$, $s\in \Theta$.  For decoding, we try to
recover $\eta_s$ fraction of $s$-input packets for each user $s$.
Precodes can be applied on the original packets of each source node so
that recovering a given fraction of the input packets of each source
node is sufficient to recover the original input packets
\cite{shokRaptor}. The precodes designed for conventional Raptor codes
can be used for our LC fountain codes. Note that the precodes usually
operate on the extension field $\ff_{q^m}$.  It is possible to use
LC fountain codes without precodes.

In this paper, we focus on three questions:
\begin{enumerate}
\item How to efficiently decode the LC fountain codes?
\item How to analyze the decoding performance?
\item How to design the degree distributions?
\end{enumerate}
The general answers to the above questions are given in
Section~\ref{sec:LC}. Before presenting the general results, 
we discuss as examples the binary LC-2 fountain code in
Section~\ref{sec:l2} and the binary LC-3 fountain
codes in Section~\ref{sec:l3}.

\subsection{Performance Bounds}
\label{sec:performance-bounds}

The coding problem described above can be regarded as coding for a
linear multiple-access channel (MAC) with $L$ inputs and one output,
where each input is a vector in $\ff_{q^m}^T$ and the output is a
sequence of linearly independent combinations of the input
vectors. The relation between the inputs and output is given by
\eqref{eq:batch}, where $H$ is only known for decoding.

Denote by $\mathcal{H}_L$ the collection of all the full-column-rank,
$L$-row matrices over $\ff_q$. $\mathcal{H}_L$ is the
set of all possible transfer matrices of the linear MAC with $L$
inputs.
Let $\mathbf{H}$ be a random matrix over $\mathcal{H}_L$.
When all the transfer matrices are independent samples of
$\mathbf{H}$, we can characterize the capacity region of the linear
MAC using the existing result on discrete memoryless MAC
\cite{cover06}. 
For an $L$-row matrix $H$ and $S \subset \{1,\ldots,L\}$, denote by
$H^S$ the submatrix of $H$ formed by the
rows indexed by $S$.
Let $R_i$ be the rate of the $i$-th input in
terms of vector per channel use. A rate tuple $(
R_1,\ldots, R_L)$ is achievable only if
\begin{equation*}
  \sum_{i\in S} R_i \leq \E[\rank(\mathbf{H}^{S})], \quad \forall S\subset \{1,\ldots,L\},
\end{equation*}
where $\mathbf{H}^S$ is the random matrix defined by
\begin{equation*}
  \Pr\{\mathbf{H}^S = H' \} = \sum_{H\in \mathcal{H}_L:H^S = H'} \Pr\{\mathbf{H}=H\}.
\end{equation*}

Further, when the
empirical distribution of the transfer matrices converges to $g$,
a rate tuple $( R_1,\ldots,  R_L)$ is achievable only if
\begin{equation*}
  \sum_{i\in S} R_i \leq \sum_{H\in \mathcal{H}_L}g(H)\rank(H^{S}), \quad \forall S\subset \{1,\ldots,L\}.
\end{equation*}
We will evaluate the performance of LC fountain codes and compare
their rate regions with the above bound.  Define
\begin{equation}\label{eq:13}
  \beta_{L} = \left(\sum_{H\in \mathcal{H}_L}g(H)\rank(H)\right).
\end{equation}
The sum rate of all inputs is upper bounded by $\beta_{L}$.

\section{LC-$2$ Fountain Codes}
\label{sec:l2}

In this section, we continue to discuss the two-user NCMA system
following Section~\ref{sec:netw-coded-mult} with the binary field as
the base field. Though they are the simplest LC fountain codes, 
LC-2 fountain codes are non-trivial and of practical
interests. 

\subsection{Parameters}
\label{sec:parameters}

When $L=2$, let $\Theta=\{\va,\vb\}$ where $\va < \vb$. We assume
$q=2$ here. As mentioned
in the introduction, for each timeslot, the nonempty
outcome of the physical layer can be grouped into four
events corresponding to four transfer matrices
\begin{equation}\label{eq:11}
  H_1=
  \begin{bmatrix}
    1 \\ 0
  \end{bmatrix},
    H_2=
  \begin{bmatrix}
    0 \\ 1
  \end{bmatrix},
  H_3=
  \begin{bmatrix}
    1 \\ 1
  \end{bmatrix},
  H_4=
  \begin{bmatrix}
    1 & 0 \\ 0 & 1
  \end{bmatrix}.
\end{equation}
 Suppose that out of the $N$ batches, transfer matrix $H_i$
occurs exactly $g(H_i)N$ times. The total number of output packets
decoded by the physical layer in $N$ timeslots is
\begin{equation*}
  n=N (g(H_1)+g(H_2)+g(H_3)+2g(H_4)) = N \beta_2,
\end{equation*}
where $\beta_2$ is defined in \eqref{eq:13}.

The output packets of an LC-2 fountain code are of two types:
\emph{clean} output packets and \emph{coupled} output packets.  A
output packet is called a \emph{clean} packet if it is an A-coded
packet or a B-coded packet. With reference to the definitions in the
introduction, the packets in $\{v_\va[i], i\in I_1\}$ and $\{v_\vb[i],
i\in I_2\}$ are clean output packets. We also simply refer to the
clean packets with respect to A and B as A-output packets and B-output
packets, respectively.  An output packet $u$ is called a
\emph{coupled} output packets if $u = v_{\va}+v_{\vb}$, where
$v_{\va}$ is an $\va$-coded packet and $v_{\vb}$ is a $\vb$-coded
packet.  The packets in $\{v_\va[i]+v_\vb[i], i\in I_3\}$ are coupled
output packets.  The numbers of A-output packets, B-output packets and
coupled output packets are $\alpha_{\va} n$, $\alpha_{\vb} n$ and
$\alpha_{\va+\vb} n$, respectively, where
\begin{IEEEeqnarray*}{rCl}
  \alpha_{\va} & = & \frac{g(H_1)+g(H_4)}{\beta_2},
  \\
  \alpha_{\vb} & = & \frac{g(H_2)+g(H_4)}{\beta_2},
  \\
  \alpha_{\va+\vb} & = & \frac{g(H_3) }{\beta_2}.
\end{IEEEeqnarray*}

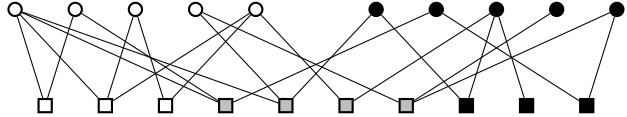
\begin{figure}
  \centering
  \begin{tikzpicture}[scale=0.8]
    \foreach \x in {1,2,...,5, 7,8,...,11}
    {
      \ifnum \x < 6
        \node[vnode] (a\x) at(\x,-0.4) {};
      \else
        \node[vnode,fill] (a\x) at(\x,-0.4) {};
      \fi
    }

    \node[cnode] at(1.5,-2) {}
    edge (a1)
    edge (a2);

    \node[cnode] at(2.5,-2) {}
    edge (a1)
    edge (a3)
    edge (a5);

    \node[cnode] at(3.5,-2) {}
    edge (a3)
    edge (a5);

    \node[cnode,fill=gray!50] at(4.5, -2) {}
    edge (a2)
    edge (a1)
    edge (a8);

    \node[cnode,fill=gray!50] at(5.5, -2) {}
    edge (a1)
    edge (a4)
    edge (a7);

    \node[cnode,fill=gray!50] at(6.5, -2) {}
    edge (a5)
    edge (a9);

    \node[cnode,fill=gray!50] at(7.5, -2) {}
    edge (a4)
    edge (a10)
    edge (a11);

    \node[cnode,fill] at(8.5,-2) {}
    edge (a7)
    edge (a9);

    \node[cnode,fill] at(9.5,-2) {}
    edge (a9);

    \node[cnode,fill] at(10.5,-2) {}
    edge (a8)
    edge (a11);
  \end{tikzpicture}
  \caption{Linearly-coupled fountain codes. The white/black circles are the
    $\va/\vb$-variable nodes, the white/black squares are the
    $\va/\vb$-check nodes, and the gray squares are the
    coupled nodes.}
  \label{fig:rateless}
\end{figure}

An LC fountain code can be
represented by a Tanner graph with the input packets as the variable
nodes and the output packets as the check nodes. We also call an input
packet a variable node and an output packet a check node henceforth.
An example of the Tanner graph is given in Fig.~\ref{fig:rateless}.

\subsection{Ordinary BP Decoding}
\label{sec:bp}

For LC-$2$ fountain codes, the \emph{(ordinary) BP decoding} of fountain
codes works well, as will be shown. In each step of the decoding
algorithm, an output packet of degree one is found, the corresponding
input packet is decoded, and the decoded input packet is substituted
into the other output packets in which it is involved. The decoding
stops when there are no more output packets of degree one.  Note that
a coupled output packet always has a degree larger than one. Hence, at
each step of the BP decoding, only an $\va$ or $\vb$-output packet of
degree one is found and decoded. Suppose that a degree-one
$\va$-output packet $u$ is found at a step of the BP decoding. Then
the $\va$-input packet embedded in $u$ can be recovered. The degrees
of the $\va$-output packets and coupled output packets embedding the
$\va$-input packet are then reduced by one. The degree reduction of
the $\va$-output packets potentially results in new degree-one
$\va$-output packets and the degree reduction of the coupled output
packets potentially results in new $\vb$-output packets, for future
steps of the BP decoding.

A check node of degree one is said to be \emph{decodable}.  There
could be multiple decodable output packets at each step of the BP
decoding. We could process the decodable output packets in different
orders. But regardless of the processing order, the algorithm will
stop with the same remaining output packets.  For example,
  the BP decoding algorithm can process all the decodable 
output packets in parallel, which is usually described as an 
  \emph{iteration based algorithm}: In each iteration, all the
  decodable output packets are found and the corresponding input
  packets are recovered, and then the recovered input packets are
  substituted into the undecodable output packets. The iteration-based
algorithm repeats the above operations until there exist no decodable
output packets.

\begin{figure}
  \centering
  \begin{tikzpicture}[scale=0.8]
    \foreach \x in {1,2,...,5, 7,8,...,11}
    {
      \ifnum \x < 6
        \node[vnode] (a\x) at(\x,-0.4) {};
      \else
        \node[vnode,fill] (a\x) at(\x,-0.4) {};
      \fi
    }
    \begin{scope}[xshift=-1cm,yshift=-2cm]
    \node[cnode] (b1) at(0,0) {}
    edge (a1)
    edge (a2);

    \node[cnode] (b2) at(1,0) {}
    edge (a1)
    edge (a3)
    edge (a5);

    \node[cnode] (b3) at(2,0) {}
    edge (a3)
    edge (a5);

    \node[cnode] (b4) at(3, 0) {}
    edge (a2)
    edge (a1);

    \node[cnode] (b5) at(4, 0) {}
    edge (a1)
    edge (a4);

    \node[cnode] (b6) at(5, 0) {}
    edge (a5);

    \node[cnode] (b7) at(6, 0) {}
    edge (a4);
    \end{scope}

    \begin{scope}[xshift=7cm,yshift=-2cm]
    \node[cnode,fill] (b8) at(0, 0) {}
    edge (a8);

    \node[cnode,fill] (b9) at(1, 0) {}
    edge (a7);

    \node[cnode,fill] (b10) at(2, 0) {}
    edge (a9);

    \node[cnode,fill] (b11) at(3, 0) {}
    edge (a10)
    edge (a11);

    \node[cnode,fill] (b12) at(4,0) {}
    edge (a7)
    edge (a9);

    \node[cnode,fill] (b13) at(5,0) {}
    edge (a9);

    \node[cnode,fill] (b14) at(6,0) {}
    edge (a8)
    edge (a11);
    \end{scope}

    \begin{scope}[xshift=-1cm,yshift=-3.5cm]
    \node[dnode] at(0,0) {}
    edge (b1);

    \node[dnode] at(1,0) {}
    edge (b2);

    \node[dnode] at(2,0) {}
    edge (b3);
    \end{scope}

    \begin{scope}[xshift=4cm,yshift=-3.5cm]
    \node[dnode,fill=gray!50] at(0,0) {}
    edge (b4)
    edge (b8);

    \node[dnode,fill=gray!50] at(1,0) {}
    edge (b5)
    edge (b9);

    \node[dnode,fill=gray!50] at(2,0) {}
    edge (b6)
    edge (b10);
    
    \node[dnode,fill=gray!50] at(3,0) {}
    edge (b7)
    edge (b11);
    \end{scope}
    
    \begin{scope}[xshift=7cm,yshift=-3.5cm]
    \node[dnode,fill] at(4,0) {}
    edge (b12);

    \node[dnode,fill] at(5,0) {}
    edge (b13);

    \node[dnode,fill] at(6,0) {}
    edge (b14);
    \end{scope}
    
  \end{tikzpicture}
  \caption{A three-layer Tanner graph for LC-2 fountain codes. The first
  layer includes the variable nodes corresponding to the input
  packets. The second layer includes the check nodes corresponding to
  the coded pakets transmitted by the source nodes. The third layer
  includes the output packets decoded by NCMA.}
  \label{fig:threelayer}
\end{figure}
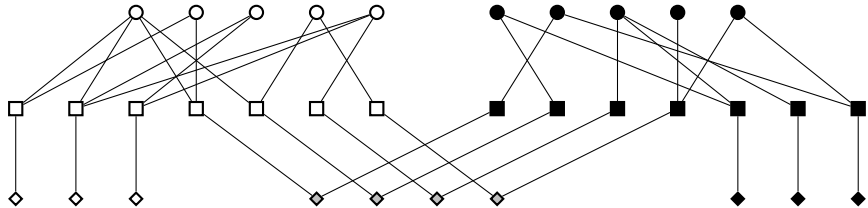

Though it is possible to analyze the BP decoding of LC-2
  fountain codes by generalizing the AND-OR tree approach introduced
  by Luby, Mitzenmacher and Shokrollahi \cite{luby98}, it would be
  difficult and/or tedious to extend this approach for general LC-$L$
  fountain codes $L>2$, where an enhanced BP decoding must be applied
  to achieved the optimal performance.  We provide an approach to
  analyze LC-$L$ fountain codes based on the existing result of LT
  codes. Here we introduce the simplified version of this
  approach for LC-2 fountain codes.  Our analysis of LC-2 fountain
codes uses the following \emph{round-based BP decoding algorithm},
which has two levels of message passing, illustrated by a three-layer
Tanner graph (see Fig.~\ref{fig:threelayer}). Each round of decoding has two
stages. In the first stage, $\va$-check nodes and $\vb$-check nodes
are decoded separately in the same manner as in conventional LT codes
until there are no decodable check nodes left. The coupled nodes are
not processed in the first stage. So the decoding in the first stage
is equivalent to decoding two LT codes in parallel. The first stage is
the message passing between the $s$-input packets and $s$-output
packets for each $s\in \Theta$, which can be analyzed using the
existing results on LT codes. In the second stage, the coupled nodes
are processed by substituting the decoded input packets. This
operation lowers the degree of coupled check nodes and may results in
new $\va$-check node and $\vb$-check node for the next round. The
second stage is the message passing between the coupled packets and
the decoded input packets, which is the essential technical part for
the analysis of LC fountain codes.

\subsection{Analysis} 

For degree distributions
$\Psi_{s}$, $s\in\Theta$, define
\begin{equation*}
  \Psi_{s}(x) = \sum_{i=1}^{D} \Psi_{s}[i] x^{i}\quad 
  \text{and}\quad 
  \Psi_{s}'(x) = \sum_{i=1}^{D}i \Psi_{s}[i] x^{i-1}.
\end{equation*}
We assume that the maximum degree $D$ does not change with the number of input packets $K_s$. This
assumption will be justified later by showing that there is a threshold on
$D$ beyond which performance will not be improved. The
following theorem tells us how many input packets are recovered for
each source node when the BP decoding stops.

\begin{theorem}
  \label{the:1}
  For each $s\in \Theta = \{\va,\vb\}$, fix $C_s>R_s > 0$.
  Consider a sequence of binary LC-$2$ fountain codes described above with $K_s/N
  \leq R_s$, $s\in \Theta$,  $N=1,2,\ldots$. 
  Define for $s,s' \in \Theta$ and $s\neq s'$,
  \begin{IEEEeqnarray*}{rCl}
    F_s(x,y) 
    & = & F_s(x,y;C_s) = \Psi_s'(x) + \frac{C_s/\beta_2}{\alpha_{s} +
      \alpha_{\va+\vb} \Psi_{s'}(y)} \ln(1-x).
  \end{IEEEeqnarray*}
  Let $z_{s}[0] = 0$ and for $i\geq 1$ let $z_{s}[i]$ be the maximum
  value of $z$ such that for any $x\in[0,\ z]$, we have
  \begin{IEEEeqnarray*}{rCl}
    F_{s}(x,z_{s'}[i-1]) \geq 0,
  \end{IEEEeqnarray*}
  where $s'\neq s$.
  The sequence $\{z_{s}[i]\}$ is increasing and upper bounded.
  Let $z_{s}^*$   be the limit of the sequence $\{z_{s}[i]\}$. 
  Then %
  with probability converging to one, as $N\rightarrow \infty$, a BP decoding algorithm
  stops with at least $z_{s}^*K_s$ $s$-input packets being decoded for all $s\in
  \Theta$.
\end{theorem}

\begin{remark}
  Consider the round-based BP decoding algorithm.  Roughly, $z_\va[i]$
  and $z_{\vb}[i]$ in the above theorem are the fractions of the decoded
  $\va$-input packets and $\vb$-input packets after the $i$-th
  round BP decoding.
\end{remark}

\begin{IEEEproof}[Sketch of the proof]
  The theorem will be proved as a special case of Theorem~\ref{the:L}
  to be presented later.
  Here we give a sketch of the proof.
  Recall an existing result of LT codes \cite{shokRaptor}. 
  Fix $ C' > R' >0$. Consider an LT code with $K$
  input packets, $n'\geq K/R'$ output packets and degree distribution
  $\Psi(x)$. If for some $0<z<1$ we have
\begin{equation*}
  \Psi'(x) + C' \ln(1-x) \geq 0,  \forall x \in [0,z],
\end{equation*}
then the code can recover at least $z K$ input packets with high
probability when $n'$ is sufficiently large.

Consider the round-based BP decoding algorithm introduced in the last
subsection. In each round, two LT codes are decoded in parallel. We
outline the analysis of the first two rounds. Taking source node $\va$
for example, in the first stage of the first round of decoding, the number of $\va$-input
packets is $K_\va$ and the number of $\va$-output packets is
$\alpha_{\va}n$. 
By the aforementioned result of LT codes, we know that with high
probability at least $z_{\va}[1]K_\va$ $\va$-input
packets can be recovered at the end of the first round when $n$ is large.

In the second stage of the first round, the decoded input packets are substituted into the coupled packets. Consider a coupled output packet $u = v_{\va} + v_{\vb}$, where $v_{\va}$ ($v_\vb$) is an
$\va$-coded ($\vb$-coded) packet. Packet $v_{\va}$ can be recovered after substitution as long as $v_{\vb}$ is a
linear combination of the decoded $\vb$-input packets. Since the set of $\vb$-input packets
embedded in $v_\vb$ is chosen uniformly, the probability that
$v_\vb$ is resolved after the first stage is at least
\begin{IEEEeqnarray*}{rCl}
  \sum_{d}\Psi_{\vb}[d] \frac{\binom{z_{\vb}[1]K_\vb}{d}}{\binom{K_\vb}{d}} & \approx & \Psi_\vb(z_{\vb}[1]).
\end{IEEEeqnarray*}
That is, the probability that $v_{\va}$ can be recovered (as an $\va$-output packet) in
the BP decoding in the second round is at least
$\Psi_{\vb}(z_{\vb}[1])$. Similarly, the probability that $v_{\vb}$ can be recovered in
the BP decoding in the second round is at least
$\Psi_{\va}(z_{\va}[1])$.

In the second round, the total number of
$\text{A}$-output packets is at least
$n[\alpha_{\va}+\alpha_{\va+\vb}\Psi_{\vb}(z_{\vb}[1])]$, and these output packets along with
the $K_\va$ $\text{A}$-input packets form an LT code. Using again the
result of LT codes, we know that at least $z_{\va}[2]K_\va$
$\text{A}$-input packets can be recovered at the end of the second
round.
\end{IEEEproof}

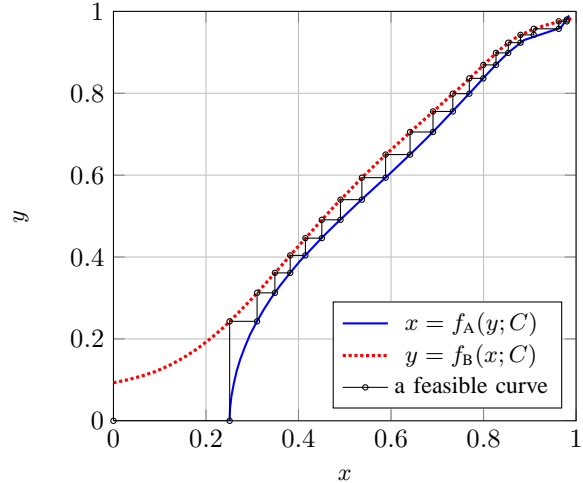
\begin{figure}
 \centering
 \begin{tikzpicture}
   \begin{axis}[
     xlabel = $x$, ylabel= $y$,
     width=220, height=200,
     xmin=0,xmax=1, ymin=0,ymax=1,
     legend pos= south east, legend style={font=\small},
     label style={font=\small},
     mark size={1.0},
     grid=both]
     \addplot[color=blue, thick, no marks] table[x=max_x1, y=y_index] {fAB.txt};
     \addplot[color=red, very thick, densely dotted, no marks]table[x=y_index, y=max_x2] {fAB.txt};
     \addplot[color=black,mark=o] table[x=xA, y=xB] {xAB.txt};
     \legend{$x=f_{\va}(y;C)$, $y=f_{\vb}(x;C)$, a feasible curve}
   \end{axis}
 \end{tikzpicture}
 \caption{Curves $x = f_{\va}(y)$ and $y = f_{\vb}(x)$ with
   $\alpha_{\va}=\alpha_{\vb}=0.25$ and $\alpha_{\va+\vb}=0.5$. The first intersection is $(0.98,0.98)$.}
 \label{fig:bp-decoding}
\end{figure}

Let us give a more explicit characterization of the limits
$(z_{\va}^*,z_{\vb}^*)$. 
Define 
\begin{IEEEeqnarray*}{rCl}
    f_{\va}(y;C_\va) & = & \max \left\{z: F_{\va}(x,y;C_\va) \geq 0,\ \forall x\in [0, z]  \right\}, \\
    f_{\vb}(x;C_\vb) & = & \max \left\{z: F_{\vb}(y,x;C_\vb) \geq
      0,\ \forall y\in [0, z] \right\}.
\end{IEEEeqnarray*}
We also write $f_\va(y;C_\va)$ and $f_\vb(x;C_\vb)$ as $f_\va(y)$ and $f_\vb(x)$, respectively, when $C_\va$ and $C_\vb$ are
implied by the context.
Both $f_\va(y)$ and $f_\vb(x)$ are increasing. The two sequences in Theorem~\ref{the:1} satisfy
$z_{\va}[i] = f_{\va}(z_{\vb}[i-1])$ and
$z_{\vb}[i] = f_{\vb}(z_{\va}[i-1])$ for $i\geq 1$.

The following lemma gives a geometric characterization of the limits
of the sequences $\{z_\va[i]\}$ and $\{z_\vb[i]\}$.

\begin{lemma}\label{lem:sd}
  The limit point $(z_{\va}^*,z_{\vb}^*)$ of the two sequences defined in
  Theorem~\ref{the:1} for LC-2 fountain codes is the \emph{first}
  intersection of the curve $x = f_{\va}(y)$ and the curve $y = f_{\vb}(x)$,
  $x,y\in [0,1]$.
\end{lemma}
\begin{IEEEproof}
  The lemma can be proved using the monotonicity of $f_\va$ and
  $f_\vb$ and is a special case of Lemma~\ref{lemma:inters}.
\end{IEEEproof}

Fig.~\ref{fig:bp-decoding} illustrates a pair of functions $f_{\va}$
and $f_{\vb}$. For a pair $(a,b)$ in the region $\{(x,y):0\leq x,
y\leq 1\}$, we say $(a,b)$ is \emph{$(C_\va,C_\vb)$-feasible} for an LC-2 fountain
code if $a \leq f_{\va}(b;C_\va)$ and $b \leq f_{\vb}(a;C_\vb)$.
A curve is \emph{$(C_\va,C_\vb)$-feasible} for an LC-2 fountain
code if every point on the curve is $(C_\va,C_\vb)$-feasible.
A point/curve is said to
be \emph{feasible} when $C_\va$ and $C_\vb$ are implied.
One property of
the feasible points is that if both $(c,d)$
and $(c,d')$ are feasible, then the vertical segment between these two
points is feasible. This is because for any $y\in [d', d]$ (assuming $d'\leq d$),
we have $y \leq d \leq f_{\vb}(c)$ and $c \leq f_{\va}(d') \leq
f_{\va}(y)$ (since $f_{\va}$ is an increasing function). The same property
holds for horizontal line segments. For example, the zig-zag curve
in Fig.~\ref{fig:bp-decoding} is a feasible curve.

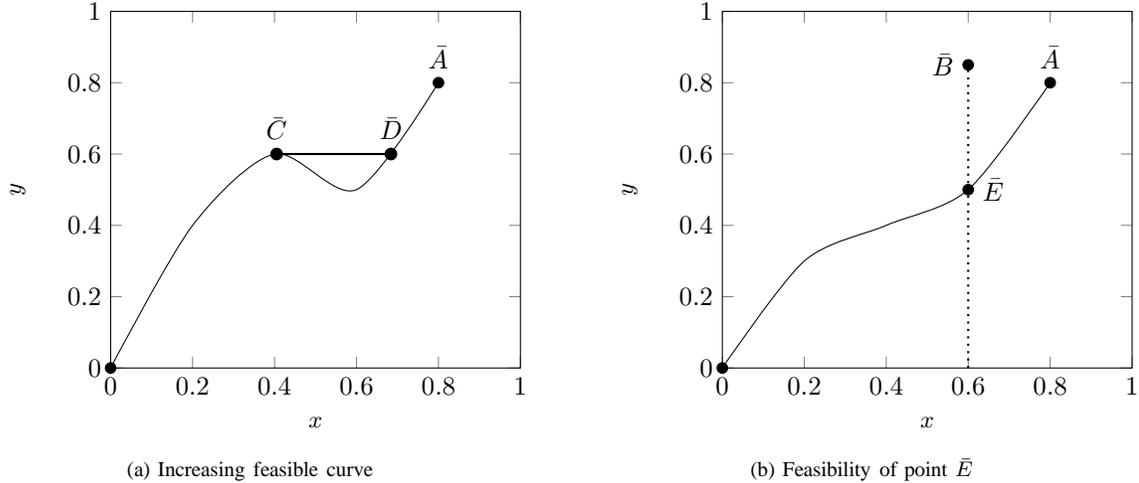
\begin{figure}
  \centering
  \begin{subfigure}[b]{0.4\textwidth}
    \centering
    \begin{tikzpicture}
      \begin{axis}[
        xlabel = $x$, ylabel= $y$,
        width=200, height=180,
        xmin=0,xmax=1, ymin=0,ymax=1,
        legend pos= south east, legend style={font=\small},
        label style={font=\small}]
        
        \addplot[no marks,smooth] coordinates 
        {(0,0) (0.2,0.4) (0.4,0.6) (0.6,0.5) (0.8,0.8)};
        
        \addplot[only marks] coordinates 
        {(0,0) (0.8,0.8)};

        \addplot[black,thick,mark=*] coordinates
        {(0.405,0.6) (0.684,0.6)};

        \node[above=2pt] at (axis cs:0.8,0.8) {$\bar{A}$};
        \node[above=2pt] at (axis cs:0.405,0.6) {$\bar{C}$};
        \node[above=2pt] at (axis cs:0.684,0.6) {$\bar{D}$};
      \end{axis}
    \end{tikzpicture}
    \caption{Increasing feasible curve}
    \label{fig:the2:1}
  \end{subfigure}
  \qquad\qquad
  \begin{subfigure}[b]{0.4\textwidth}
    \centering
    \begin{tikzpicture}
      \begin{axis}[
        xlabel = $x$, ylabel= $y$,
        width=200, height=180,
        xmin=0,xmax=1, ymin=0,ymax=1,
        legend pos= south east, legend style={font=\small},
        label style={font=\small}]
        
        \addplot[no marks,smooth] coordinates 
        {(0,0) (0.2,0.3) (0.4,0.4) (0.6,0.5) (0.8,0.8)};

        \addplot[no marks, thick, dotted] coordinates
        {(0.6,0) (0.6,0.85)};
        
        \addplot[only marks] coordinates 
        {(0,0) (0.8,0.8) (0.6,0.85) (0.6,0.5)};

        \node[above=2pt] at (axis cs:0.8,0.8) {$\bar{A}$};
        \node[left=2pt] at (axis cs:0.6,0.85) {$\bar{B}$};
        \node[right=2pt] at (axis cs:0.6,0.5) {$\bar{E}$};

      \end{axis}
    \end{tikzpicture}
    \caption{Feasibility of point $\bar{E}$}\label{fig:the2:2}
  \end{subfigure}

  \caption{An illustration of the proof of Theorem~\ref{the:2}.}
  \label{fig:the2}
\end{figure}

\begin{theorem}
  \label{the:2}
  For each $s\in\Theta=\{\va,\vb\}$, fix 
  $C_s > R_s > 0$. Consider a sequence of binary LC-2
  fountain codes with $N=1,2\ldots$, where $K_s/N \leq
   R_s$ for $s\in\Theta$. For any pair $(a_\va,a_\vb)$, if there exists a
  $(C_\va,C_\vb)$-feasible continuous curve $(x(t),y(t))$ between the origin
  and
  $(a_\va,a_\vb)$, then i) a BP decoding algorithm will stop with at least
  $a_sK_s$ $s$-input packets being decoded for all $s\in\Theta$ with
  probability converging to one when $N\rightarrow \infty$, and ii)
  there exists an \emph{increasing}, continuous and $(C_\va,C_\vb)$-feasible curve $(\tilde x(t),\tilde y(t))$ between the origin and $(a_\va,a_\vb)$.
\end{theorem}
\begin{IEEEproof}[Sketch of the proof]
  The theorem will be proved as a special case of Theorem~\ref{the:L2}.
  Here we give a sketch of the proof.
Fig.~\ref{fig:the2} illustrates the main ideas, in which the point $(a_\va,a_\vb)$ is labeled by $\bar{A}$. We first show the second claim. Suppose there exists a feasible curve from the origin to point $\bar{A}$, which is not increasing, e.g., the thin solid curve in Fig.~\ref{fig:the2:1}. Point $\bar{C}$ is a local maximum of
the curve and point $\bar{D}$ is also on the curve which share the same
$y$-coordinate as point $\bar{C}$. We can replace the part of the curve between
points $\bar{C}$ and $\bar D$ by the line segment (the thick solid line segment in
the figure) between points $\bar C$ and $\bar D$. The new curve is 
increasing. The points on the line segment between points $\bar C$ and $\bar D$ are
feasible since both $\bar C$ and $\bar D$ are feasible. The second claim in the
theorem can be proved by repeating the above procedure. 

It is sufficient to prove the first claim for increasing
curve $(x(t),y(t))$. Fix $C_\va'$ and $C_\vb'$ such that
$R_\va<C_\va'<C_\va$ and $R_\vb<C_\vb'<C_\vb$.
Denote by $\bar{B}=(b_\va,b_\vb)$ the first intersection
of curves $x=f_\va(y;C_\va')$ and $y=f_\vb(x;C_\vb')$. If both $b_\va \geq a_\va$
and $b_\vb\geq a_\vb$, the first claim holds by Lemma~\ref{lem:sd} and
Theorem~\ref{the:1}. We then show by contradiction that it is not
possible that either $b_\va < a_\va$ or $b_\vb < a_\vb$. Suppose
$b_\va < a_\va$ and $b_\vb \geq a_\vb$ as illustrated in Fig. \ref{fig:the2:2}. Consider the
point $\bar E=(b_\va,e_\vb)$ on the curve $(x(t),y(t))$, where $e_\vb
\leq a_\vb \leq b_\vb$. 
The contradiction is that $\bar E$ is not
$(C_\va,C_\vb)$-feasible since
\begin{equation*}
  b_\va  =  f_\va(b_\vb,C') \geq  f_\va(e_\vb,C')  >  f_{\va}(e_\vb,C),
\end{equation*}
where the inequalities follow from the monotonicity of the function $f_\va$.
\end{IEEEproof}

\subsection{Optimizations}

Given the parameters $\alpha_\va$, $\alpha_\vb$ and
$\alpha_{\va+\vb}$, we want to design a binary LC-2
fountain codes such that at least $\eta_\va$ fraction of $\va$-input
packets and $\eta_\vb$ fraction of $\vb$-input packets can be decoded
by BP decoding. By Theorem~\ref{the:2}, a rate pair $(\eta_\va C_\va, \eta_\vb C_\vb)$ is \emph{achievable}
by BP decoding if there exists a $(C_\va,C_\vb)$-feasible
curve between the origin and $(\eta_\va,\eta_\vb)$. Theorem~\ref{the:2} also enables us to consider only
the increasing curves from the origin to $(\eta_\va,\eta_\vb)$.

By definition, a point $(\hat{x},\hat{y})$ is $(C_\va,C_\vb)$-feasible
if $\hat{x} \leq f_\va(\hat{y};C_\va)$ and $\hat{y}\leq
f_\vb(\hat{x};C_\vb)$, which are equivalent to
\begin{IEEEeqnarray*}{rCl}
  F_\va(x,\hat{y};C_\va) & \geq & 0, \quad \forall x \in [0,\hat{x}], \\
  F_\vb(y,\hat{x};C_\vb) & \geq & 0, \quad \forall y \in [0,\hat{y}],
\end{IEEEeqnarray*}
that is,
\begin{IEEEeqnarray}{rCl}
  \left[\alpha_{\va} + \alpha_{\va+\vb} \label{eq:con1}
    \Psi_{\vb}(\hat{y})\right] \Psi_\va'(x) + {C_\va/\beta_2} \ln(1-x) & \geq & 0, \quad \forall x \in [0,\hat{x}], \\
  \left[\alpha_{\vb} + \alpha_{\va+\vb} \label{eq:con2}
    \Psi_{\va}(\hat{x})\right]\Psi_\vb'(y) + {C_\vb/\beta_2} \ln(1-y) & \geq & 0, \quad \forall y \in [0,\hat{y}].
\end{IEEEeqnarray}
We only evaluate the zig-zag type of curves (see
Fig.~\ref{fig:bp-decoding} for an example). Fix a positive integer
$t_{\max}$ and two sequences of real numbers $x_t,y_t$,
$t=0,1,\ldots,t_{\max}$ with 
\begin{IEEEeqnarray*}{l}
0=x_0\leq x_1 \leq \ldots \leq
x_{t_{\max}} = \eta_\va, \\
0=y_0\leq y_1 \leq \ldots \leq
y_{t_{\max}} = \eta_\vb.
\end{IEEEeqnarray*}
The curve formed by line segments $(x_t,y_t) - (x_{t+1},y_t) - (x_{t+1},
y_{t+1})$, $t=0, 1,\ldots, t_{\max}-1$ is an increasing zig-zag curve from the origin to
$(\eta_\va,\eta_\vb)$. As explained before, the vertical (horizontal) line segment between two feasible points is feasible. So we
only need to check the feasibility of the points
\begin{equation}
  \label{eq:2}
  (x_0,y_0),
(x_1,y_0), (x_1,y_1), (x_2,y_1), \ldots, (x_{t_{\max}},y_{t_{\max}}).
\end{equation}
We do not lose optimality since all increasing curves can be
approximated closely by such zig-zag curves when $t_{\max}$ is
sufficiently large.

Now we are ready to introduce the optimization problems for binary
LC-2 fountain codes. Since we have a pair of coding rates,
we may fix one and maximize the other or maximize the sum rate. 
Fix $t_{\max}$, $C_\vb$, $\eta_{\va}$ and $\eta_{\vb}$.
The following optimization problem maximizes the achievable rate of 
source node $\va$ for a given rate of source node $\vb$:
\begin{equation}
  \label{eq:op1}
    \begin{IEEEeqnarraybox}[][c]{r.l}
      \max & \eta_\va\theta_\va\beta_2 \\
      \text{s.t.} & 
        x_0 = 0, y_0 = 0, x_{t_{\max}} = \eta_{\va}, y_{t_{\max}} =
       \eta_{\vb},\\
       & \forall t = 1, \ldots, t_{\max}, \quad x_{t} \geq x_{t-1}, y_{t} \geq y_{t-1}, \\
       & \left[\alpha_{\va} + \alpha_{\va+\vb}
         \Psi_{\vb}\left(y_{t-1}\right)\right]\Psi_{\va}'(x) +
       \theta_{\va} \ln(1-x)\geq 0, 
        \quad \forall x \in (x_{t-1}, x_t],\\
       & \left[\alpha_{\vb} + \alpha_{\va+\vb}
         \Psi_{\va}\left(x_{t}\right)\right]\Psi_{\vb}'(y) + 
       C_\vb/\beta_2 \ln(1-y)\geq 0, 
        \quad \forall y \in (y_{t-1}, y_t],
    \end{IEEEeqnarraybox}
\end{equation}
where the variables of the optimization are $\theta_\va$, $x_t$,
$y_t$, $t=1,\ldots,t_{\max}$, $\Psi_\va$ and
$\Psi_\vb$. Note that in the above optimization, we do not require the
inequalities in the last two lines to be satisfied for $x$ or $y$
starting from zero as in \eqref{eq:con1} and \eqref{eq:con2}. But the
last two lines can still guarantee that the points in \eqref{eq:2} are all
feasible due to the following property. Suppose that for $i=1, \ldots, t$ we have
\begin{equation*}
  \left[\alpha_{\vb} + \alpha_{\va+\vb}
         \Psi_{\va}\left(x_{i}\right)\right]\Psi_{\vb}'(y) + 
       C_\vb/\beta_2 \ln(1-y)\geq 0, 
        \quad \forall y \in (y_{i-1}, y_i].
\end{equation*}
Due to the monotonic property of $\Psi_{\va}(x)$ and $x_t\geq x_i$ for $i<t$, we have for $i=1, \ldots, t$
\begin{equation*}
  \left[\alpha_{\vb} + \alpha_{\va+\vb}
         \Psi_{\va}\left(x_{t}\right)\right]\Psi_{\vb}'(y) + 
       C_\vb/\beta_2 \ln(1-y)\geq 0, 
        \quad \forall y \in (y_{i-1}, y_i].
\end{equation*}
Combining the $t$ equalities, we have 
\begin{equation*}
  \left[\alpha_{\vb} + \alpha_{\va+\vb}
         \Psi_{\va}\left(x_{t}\right)\right]\Psi_{\vb}'(y) + 
       C_\vb/\beta_2 \ln(1-y)\geq 0, 
        \quad \forall y \in (0, y_t].
\end{equation*}
Similarly, the second last line in the above optimization implies
\begin{equation*}
  \left[\alpha_{\va} + \alpha_{\va+\vb}
         \Psi_{\vb}\left(y_{t-1}\right)\right]\Psi_{\va}'(x) +
       \theta_{\va} \ln(1-x)\geq 0, 
        \quad \forall x \in (0, x_t].
\end{equation*}
We can also write an optimization to maximize the rate of the source
node $\vb$.

For given 
$t_{\max}$, $\eta_{\va}$ and $\eta_{\vb}$, we can maximize the sum
rate of both source nodes as follows: 
\begin{equation}
  \label{eq:op2}
    \begin{IEEEeqnarraybox}[][c]{r.l}
      \max & \beta_2(\eta_\va\theta_\va+\eta_\vb\theta_\vb) \\
      \text{s.t.} & 
        x_0 = 0, y_0 = 0, x_{t_{\max}} = \eta_{\va}, y_{t_{\max}} =
       \eta_{\vb},\\
       & \forall t = 1, \ldots, t_{\max}, \quad x_{t} \geq x_{t-1}, y_{t} \geq y_{t-1}, \\
       & \left[\alpha_{\va} + \alpha_{\va+\vb}
         \Psi_{\vb}\left(y_{t-1}\right)\right]\Psi_{\va}'(x) +
       \theta_{\va} \ln(1-x)\geq 0, 
        \quad \forall x \in (x_{t-1}, x_t],\\
       & \left[\alpha_{\vb} + \alpha_{\va+\vb}
         \Psi_{\va}\left(x_{t}\right)\right]\Psi_{\vb}'(y) + 
       \theta_\vb \ln(1-y)\geq 0, 
        \quad \forall y \in (y_{t-1}, y_t],
    \end{IEEEeqnarraybox}
\end{equation}
where the variables of the optimization are $\theta_\va$,
$\theta_\vb$, $x_t$, $y_t$, $t=1,\ldots,t_{\max}$, degree
distributions $\Psi_\va$ and $\Psi_\vb$.

The maximum degree $D$ can be similarly bounded as for fountain codes.

\begin{lemma}\label{lem:4}
  Consider optimizations \eqref{eq:op1} and  \eqref{eq:op2}. %
  For $s\in \{\va,\vb\}$, using degrees larger than $\lceil
  1/(1-\eta_{s})\rceil -1$ for $\Psi_{s}$ does not give better optimal
  values.
\end{lemma}
\begin{IEEEproof}
  We use problem \eqref{eq:op2} as an example to prove the lemma.
  Consider an integer $\Delta$ such that $1 - \eta_{\va} \geq
  \frac{1}{\Delta+1}$. 
  Let $\Psi_{\va}$ be a degree distribution with
  $\sum_{d>\Delta} \Psi_{\va}[d] >0$. Construct a new degree
  distribution   $\tilde \Psi_{\va}$ with
  $\tilde \Psi_{\va}[d] = \Psi_{\va}[d]$ for $d<\Delta$, 
  $\tilde \Psi_{\va}[\Delta]  =  \sum_{d\geq \Delta} \Psi_{\va}[d]$ and
  $\tilde \Psi_{\va}[d]  =  0$ for $d>\Delta$.
We have $\tilde\Psi_{\va}(x) - \Psi_{\va}(x) =
\sum_{d>\Delta}\Psi_{\va}[d] (x^{\Delta} - x^{d}) > 0$ and 
\begin{IEEEeqnarray*}{rCl}
  \tilde\Psi_{\va}'(x) - \Psi_{\va}'(x)
  & = & 
  \sum_{d> \Delta} \Psi_{\va}[d] (\Delta x^{\Delta-1} - d x^{d-1}).
\end{IEEEeqnarray*}
Since for $d\geq \Delta$
\begin{equation*}
  \frac{(d+1)x^{d}}{d x^{d-1}} = \frac{d+1}{d} x \leq 
  \frac{d+1}{d} \eta_{\va}
   \leq  \frac{\Delta+1}{\Delta} \eta 
   =  1,
\end{equation*}
we have $\tilde\Psi_{\va}'(x) \geq \Psi_{\va}'(x)$. 
Thus, $\tilde \Psi_{\va}$ does not give worse optimal value than
$\Psi_{\va}$. The part of the lemma for $\Psi_{\vb}$ can be similarly
proved.
\end{IEEEproof}

\subsection{Achievable Rates}
\label{sec:numer-eval}

\begin{table}
  \centering
  \caption{Achievable rates of binary LC-2 fountain codes for
    $\eta_\va = \eta_\vb = 0.98$. In both \eqref{eq:op1} and
    \eqref{eq:op2}, the objective functions are modified by removing
    $\beta_2$. $\hat R_\va/\beta_2$ is obtained
    by solving \eqref{eq:op1} with $C_\vb/\beta_2 =
    \alpha_\vb/\eta_\vb$, and
    $\hat R_{\text{sum}}/\beta_2$ is obtained
    by solving~\eqref{eq:op2}.}
  \label{tab:1}
  \begin{tabular}{c|c|c|c|c}
    \hline
    $\alpha_{\va+\vb}$ & $\alpha_{\va}$ &
    $\alpha_{\vb}$ & $\hat R_\va/\beta_2$  & $\hat R_{\text{sum}}/\beta_2$ \\
    \hline\hline
    0.05 & 0.475 & 0.475 & 0.5135 & 0.9879 \\
    \hline
    \multirow{2}{*}{0.25} & 0.375 & 0.375 & 0.5962 & 0.9797 \\
    & 0.45 & 0.3 & 0.6701 & 0.9823 \\
    \hline
    \multirow{3}{*}{0.5} & 0.25 & 0.25 & 0.7022 & 0.9617 \\
    & 0.375 & 0.125 & 0.8292 & 0.9724  \\
    & 0.45 & 0.05 & 0.9090 & 0.9616 \\
    \hline
    \multirow{3}{*}{0.75} &  0.125 & 0.125 & 0.8137 & 0.9510 \\
    & 0.1875 & 0.0625 & 0.8854 & 0.9571 \\
    & 0.225 & 0.025 & 0.9359 & 0.9589 \\
    \hline
    0.95 & 0.025 & 0.025 & 0.9317 & 0.9496 \\
    \hline
  \end{tabular}
\end{table}

Given the distribution $g$ of the transfer matrix, we know from
Section~\ref{sec:performance-bounds} that a rate
pair $( R_\va, R_\vb)$ is achievable only if
\begin{IEEEeqnarray*}{rCl}
   R_\va & \leq & \sum_i g(H_i)\rank(H_i^{\{\va\}}) = g(H_1) + g(H_3) +
  g(H_4) = \beta_2(\alpha_\va+\alpha_{\va+\vb}), \\
   R_\vb & \leq & \sum_i g(H_i)\rank(H_i^{\{\vb\}}) = g(H_2) + g(H_3) +
  g(H_4) = \beta_2(\alpha_\vb+\alpha_{\va+\vb}), \\
   R_\va +  R_\vb & \leq & \sum_i g(H_i) \rank(H_i) = \beta_2.
\end{IEEEeqnarray*}

Instead of specifying a value of $\beta_2$, we remove $\beta_2$ from
the objective functions of both \eqref{eq:op1} and \eqref{eq:op2} so
that the optimal values are the normalized (sum) rates.   The
  best numerical results obtained by evaluating the modified
  optimization \eqref{eq:op2} are listed in Table~\ref{tab:1}, where we
  can see that the normalized achievable sum rates are all close
  $1$, the upper bound. One of the vertex
of the above region is $R_\va = \beta_2(\alpha_\va+\alpha_{\va+\vb})$
and $R_\vb = \beta_2\alpha_\vb$. We evaluate \eqref{eq:op1} with
$C_\vb/\beta_2 = \alpha_\vb/\eta_\vb$. From
  Table~\ref{tab:1}, readers can verify that the normalized achievable
  rates of user $\va$ are all close to the corresponding values of
  $\alpha_\va+\alpha_{\va+\vb}$. Note that for the values obtained in
Table~\ref{tab:1}, $\beta_2$ can be any value in the range $(0,2)$.

 The optimizations \eqref{eq:op1} and \eqref{eq:op2} are
non-convex and hence we may not obtain the globally optimal values.
We discuss in the appendix how to solve these optimizations.
Nevertheless, the numerical results show that the obtained suboptimal
rates are all very close to the bound we provided above. Since the
values may not be globally optimal, for each row it is possible that
the value of $\alpha_\vb$ plus the value in the second last
column is larger than the value in the last column.

\section{LC-$3$ Fountain Codes}
\label{sec:l3}

Our discussion of LC-2 function codes can be generalized to LC-$L$
with $L > 2$. However, the generalization involves new features absent
in the LC-2 case.  In this section, 
we use the LC-3 fountain codes to illustrate the implications of these
new features for the design and analysis of general LC-$L$ fountain
codes. 

\subsection{Batches}
\label{sec:batches}

For $L=3$, let $\Theta=\{\va,\vb,\text{C}\}$, where $\va<\vb<\vc$. We
assume $q=2$ here.  Compared with LC-2 fountain codes, we have a new
type of coupled packet $v_\va+v_\vb+v_\vc$ embedded with three (rather
than just two) coded packets, where $v_s$, $s\in\Theta$ is transmitted
by source node $s$.  We say an output packet of a batch is
\emph{autonomous} if none of the coded packets embedded in it is
embedded in other output packets of the batch. For example, if the
physical layer decodes $v_\va$ and $v_\va +v_\vb+v_\vc$, we get two
non-autonomous output packets. But we can transform them into autonomous
output packets by reducing $v_\va +v_\vb+v_\vc$ to $v_\vb+v_\vc$.  On
the other hand, if the physical layer decodes $v_\va+v_\vb$ and
$v_\vb+v_\vc$, we cannot transform them into autonomous output
packets.

For each timeslot, if the physical layer decodes only one packet, the
packet is autonomous. If the physical layer decodes three linearly
independent packets, after linear transformation, this is equivalent
to obtaining three autonomous output packets $v_\va$, $v_\vb$ and
$v_\vc$.  If the physical layer decodes two linearly independent
packets, it is possible to have non-autonomous output packets as seen
in the above example. For an LC-3 fountain code, all
non-autonomous output packets can be put into the form $\{v_\va+v_\vc,
v_\vb+v_\vc\}$ after linear transformation. We will see that
to achieve optimal performance, non-autonomous output packets
should be handled in a different way from how autonomous output packets
are handled.

The combined decoding outcomes of the physical
layer, after proper linear transformation, can be categorized into the following eight cases:
\begin{enumerate}
\item Only $v_s$ is decoded, where $s\in\Theta$. The corresponding
  transfer matrix is one of the following:
    \begin{equation*}
    H_1=
    \begin{bmatrix}
      1 \\ 0 \\ 0
    \end{bmatrix},
    H_2=
    \begin{bmatrix}
      0 \\ 1 \\ 0
    \end{bmatrix},
    H_3=
    \begin{bmatrix}
      0 \\ 0 \\ 1
    \end{bmatrix}.
  \end{equation*}
\item Only $v_s$ and $v_{s'}$ are
  decoded, where $s < s'\in\Theta$. The corresponding transfer
  matrix is one of the following:
  \begin{equation*}
    H_4=
    \begin{bmatrix}
      1 & 0 \\ 0 & 1 \\ 0 & 0
    \end{bmatrix},
    H_5=
    \begin{bmatrix}
      1 & 0 \\ 0 & 0 \\ 0 & 1
    \end{bmatrix},
    H_6=
    \begin{bmatrix}
      0 & 0 \\ 1 & 0 \\ 0 & 1 
    \end{bmatrix}.
  \end{equation*}
\item All the three packets $v_{\va}$,
$v_\vb$ and $v_\vc$ are decoded. The corresponding transfer matrix is 
\begin{equation*}
      H_7=
    \begin{bmatrix}
      1 & 0 & 0 \\ 0 & 1 & 0 \\ 0 & 0 & 1 
    \end{bmatrix}.
\end{equation*}
\item Only $v_s+v_{s'}$ is decoded, where $s < s'\in\Theta$. The
  corresponding transfer matrix is one of the following:
  \begin{equation*}
     H_8=
    \begin{bmatrix}
      1 \\ 1 \\ 0
    \end{bmatrix},
    H_9=
    \begin{bmatrix}
      1 \\ 0 \\ 1
    \end{bmatrix},
    H_{10}=
    \begin{bmatrix}
      0 \\ 1 \\ 1
    \end{bmatrix}.
  \end{equation*}
\item Only $v_\va+v_\vb+v_\vc$ is decoded. The corresponding transfer matrix is
  \begin{equation*}
        H_{11}=
    \begin{bmatrix}
      1 \\ 1 \\ 1
    \end{bmatrix}.
  \end{equation*}

\item Only $v_s+v_{s'}$ and $v_{s''}$
  are decoded, where $s \neq s'\neq s''\in\Theta$ and $s<s'$. The
  corresponding transfer matrix is one of the following:
  \begin{equation*}
    H_{12}=
    \begin{bmatrix}
      1 & 0 \\ 0 & 1 \\ 1 & 0
    \end{bmatrix},
    H_{13}=
    \begin{bmatrix}
      1 & 0 \\ 0 & 1 \\ 0 & 1
    \end{bmatrix},
    H_{14}=
    \begin{bmatrix}
      1 & 0 \\ 1 & 0 \\ 0 & 1 
    \end{bmatrix}.
  \end{equation*}

\item Two non-autonomous output packets are decoded. The
  corresponding transfer matrix is one of the following:
  \begin{equation*}
    H_{15}=
    \begin{bmatrix}
      1 & 0 \\ 0 & 1 \\ 1 & 1
    \end{bmatrix},
    H_{16}=
    \begin{bmatrix}
      1 & 1 \\ 0 & 1 \\ 1 & 0
    \end{bmatrix},
    H_{17}=
    \begin{bmatrix}
      1 & 0 \\ 1 & 1 \\ 0 & 1
    \end{bmatrix}.
  \end{equation*}
\item Nothing is decoded.
\end{enumerate}
Suppose that the number of batches with the transfer matrix $H_i$
occurring is exactly $g(H_i)N$.  The total number of output packets is
$n=\beta_3 N$, where $\beta_3$ is defined in \eqref{eq:13}.

An autonomous output packet of the form $\sum_{s\in S} v_s$ for
certain $S\subset \Theta$ is called an \emph{$S$-output packet}. 
Define for LC-3 fountain codes
\begin{IEEEeqnarray*}{rCl}
  \alpha_{\va} & = & %
  \frac{g(H_1)+g(H_4)+g(H_5)+g(H_7)+g(H_{13})}{\beta_3},
  \\
  \alpha_{\vb} & = &
  \frac{g(H_2)+g(H_4)+g(H_6)+g(H_7)+g(H_{12})}{\beta_3},
  \\
  \alpha_{\vc} & = &
  \frac{g(H_3)+g(H_5)+g(H_6)+g(H_7)+g(H_{14})}{\beta_3}, \\
  \alpha_{\va+\vb} & = & 
  \frac{g(H_8)+g(H_{14})}{\beta_3}, \\
  \alpha_{\va+\vc} & = & 
  \frac{g(H_9)+g(H_{12})}{\beta_3}, \\
  \alpha_{\vb+\vc} & = & 
  \frac{g(H_{10})+g(H_{13})}{\beta_3}, \\
  \alpha_{\va+\vb+\vc} & = & 
  \frac{g(H_{11})}{\beta_3}, \\
  \bar \alpha_\va & = & \frac{g(H_{16})}{\beta_3},\\
  \bar \alpha_\vb & = & \frac{g(H_{17})}{\beta_3},\\
  \bar \alpha_\vc & = &  %
  \frac{g(H_{15})}{\beta_3}.
\end{IEEEeqnarray*}
For $s\neq s'\neq s''$, we also write $\alpha_s=\alpha_{\{s\}}$,
$\alpha_{s+s'} = \alpha_{\{s,s'\}}$ and
$\alpha_{s+s'+s''}=\alpha_{\{s,s',s''\}}$.  We have
\begin{equation*}
  \sum_{S\subset\Theta:|S|\geq 1} \alpha_{S} + 2 \sum_{s\in\Theta}\bar \alpha_s = 1.
\end{equation*}
For each $S\subset \Theta$
and $S\neq \emptyset$, the number of (autonomous) $S$-output packets
is $\alpha_{S}n$.  When $S =\{s\}$, an $S$-output packet is an
$s$-output packet. Totally, we have $n \sum_{S\subset\Theta:|S|\geq 1}
\alpha_{S}$ autonomous output packets. Let
\begin{equation*}
  \bar \alpha = \bar \alpha_\va+ \bar \alpha_\vb + \bar \alpha_\vc.
\end{equation*}
The remaining $n
(1-\sum_{S\subset\Theta:|S|\geq 1} \alpha_{S}) = 2 n \bar \alpha$
output packets are non-autonomous output packets contained in $n\bar\alpha = N
[g(H_{15}) +g(H_{16})+g(H_{17})]$ batches.

\subsection{Batched BP Decoding}

The ordinary BP decoding of fountain codes can be used to decode
LC-3 fountain codes. But as we will show in the next example, we can
improve the decoding performance by exploiting the batch  structure of
the non-autonomous output packets in the decoding process.

\begin{figure}
  \centering
  \begin{tikzpicture}[scale=0.8]
    \foreach \x in {1,2,3,4,6,7,8,9,11,12,13,14}
    {
        \node[vnode] (a\x) at(\x,-0.4) {};
    }

    \begin{scope}[xshift=0.5cm,yshift=-2cm]
    \node[cnode] (b1) at(0,0) {}
    edge (a1)
    edge (a2);

    \node[cnode] (b2) at(1,0) {}
    edge (a1)
    edge (a3)
    edge (a4);

    \node[cnode] (b3) at(2,0) {}
    edge (a3)
    edge (a4);

    \node[cnode,label=below:$v_\va$] (b4) at(3, 0) {}
    edge (a2)
    edge (a1);

    \node[cnode] (b5) at(4, 0) {}
    edge (a1) 
    edge (a4);
    \end{scope}
    
    \begin{scope}[xshift=5.5cm,yshift=-2cm]
    \node[cnode,label=below:$v_\vb$] (b6) at(0, 0) {}
    edge (a7)
    edge (a6);

    \node[cnode] (b7) at(1, 0) {}
    edge (a8)
    edge (a6);
    
    \node[cnode] (b8) at(2, 0) {}
    edge (a6)
    edge (a9)
    edge (a8);

    \node[cnode] (b9) at(3, 0) {}
    edge (a8)
    edge (a9)
    edge (a7);

    \node[cnode] (b10) at(4, 0) {}
    edge (a7)
    edge (a9);
    \end{scope}

    \begin{scope}[xshift=10.5cm,yshift=-2cm]
    \node[cnode,label=below:$v_\vc$] (b11) at(0, 0) {}
    edge (a11);

    \node[cnode] (b12) at(1,0) {}
    edge (a12)
    edge (a13);

    \node[cnode] (b13) at(2,0) {}
    edge (a12)
    edge (a14);

    \node[cnode] (b14) at(3,0) {}
    edge (a12)
    edge (a11);

    \node[cnode] (b15) at(4,0) {}
    edge (a14)
    edge (a11);
    \end{scope}

    \begin{scope}[xshift=0.5cm,yshift=-3.5cm]
    \node[dnode] at(0,0) {}
    edge (b1);

    \node[dnode] at(1,0) {}
    edge (b2);

    \node[dnode] at(2,0) {}
    edge (b3);
    \end{scope}

    \begin{scope}[xshift=4.5cm,yshift=-3.5cm]
    \node[dnode,fill=gray!50,label=below:$u_1$] at(0,0) {}
    edge (b4)
    edge (b6);

    \node[dnode,fill=gray!50] at(1,0) {}
    edge (b5)
    edge (b7);

    \node[dnode,fill=gray!50,label=below:$u_2$] at(2,0) {}
    edge (b6)
    edge (b11);
    
    \node[dnode,fill=gray!50] at(6,0) {}
    edge (b10)
    edge (b12);
    \end{scope}
    
    \begin{scope}[xshift=7.5cm,yshift=-3.5cm]
    \node[dnode] at(0,0) {}
    edge (b8);

    \node[dnode] at(1,0) {}
    edge (b9);
    \end{scope}
    
    \begin{scope}[xshift=12.5cm,yshift=-3.5cm]
    \node[dnode] at(0,0) {}
    edge (b13);

    \node[dnode] at(1,0) {}
    edge (b14);

    \node[dnode] at(2,0) {}
    edge (b15);
    \end{scope}
    
  \end{tikzpicture}
  \caption{A three-layer Tanner graph for LC-3 fountain codes. The first
  layer includes the variable nodes corresponding to the input
  packets. The second layer includes the check nodes corresponding to
  the coded pakets transmitted by the source nodes. The third layer
  includes the output packets decoded by NCMA. In this graph, $u_1$
  and $u_2$ forms a batch with two non-autonomous packets.}
  \label{fig:batch}
\end{figure}
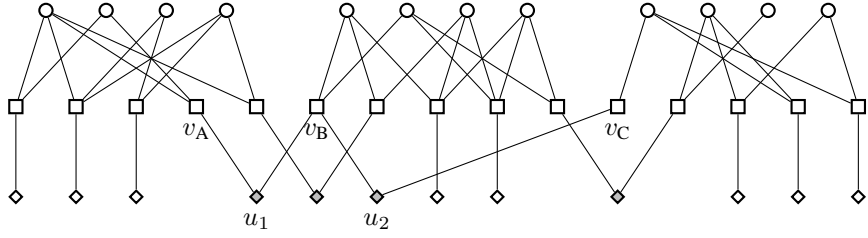

Consider a batch of two non-autonomous output packets $u_1 =
v_\va+v_\vb$ and $u_2 = v_\vb+v_\vc$ (see the illustration in
Fig.~\ref{fig:batch}).  Suppose that when the ordinary BP decoding
stops, packet $v_\va$ is a linear combination of the already-decoded
$\va$-input packets, packet $v_\vb$ has a degree larger than one, and
packet $v_\vc$ has degree one. The ordinary BP decoding substitutes
the already-decoded $\va$-input packets in $u_1$ and recovers
$v_\vb$. But since only already-decoded input packets can be
substituted, the ordinary BP decoding does not substitute $v_\vb$ into
$u_2$ to recover $v_\vc$, and hence the BP decoding cannot be
resumed. However, if we allow joint processing of $u_1$ and $u_2$, we
can substitute $v_\vb$ into $u_2$ to obtain $v_\vc$ and hence the BP
decoding can be resumed since $v_\vc$ has degree one.

Motivated by the above example, we propose the \emph{batched BP
  decoding} for LC-3 codes. Recall that only batches with transfer
matrices $H_{15}, H_{16}$ and $H_{17}$ have non-autonomous output
packets. The batched BP
decoding is the same as the ordinary BP decoding except that it also
solves the linear systems of equations (at the second stage of each round):
\begin{equation}\label{eq:8}
  [u_1, u_2]
  =
  [v_\va, v_\vb, v_\vc]
  H_{15}, 
\end{equation}
where $u_1$ and $u_2$ are the two output packets of the batch. Note
that for batches with transfer matrices $H_{16}$ and $H_{17}$, the
associated linear systems are equivalent to \eqref{eq:8}. When
any one of $v_\va, v_\vb$ or $v_\vc$ is the linear combination of the
already-decoded input packets, the batched BP decoding solves
\eqref{eq:8} to resolve the value of the other two.

\subsection{Analysis}

The following theorem tells us how many
input packets are recovered for each source node when the \emph{ordinary} BP
decoding stops for binary LC-3 fountain codes.

\begin{theorem}
  \label{the:l3o}
  For each $s\in \Theta = \{\va,\vb,\vc\}$, fix $C_s>R_s>0$ and
  consider a sequence of binary LC-$3$ fountain codes described above
  with $K_s/N \leq R_s$, $s\in \Theta$, $N=1,2,\ldots$. 
  For $s\neq s' \neq s''\in \Theta$, define
  \begin{IEEEeqnarray*}{rCl}
    F_s^o(x,x',x'')
    & = & \Psi_s'(x) + \frac{C_s/\beta_3}{\alpha_{s} + \lambda_1(s) + \lambda_2^o(s)}\ln(1-x),
  \end{IEEEeqnarray*}
  where
  \begin{IEEEeqnarray*}{rCl}
    \lambda_1(s) & = & 
    \alpha_{s+s'} \Psi_{s'}(x') + \alpha_{s+s''}\Psi_{s''}(x'') +
    \alpha_{s+s'+s''}\Psi_{s'}(x') \Psi_{s''}(x''),\\
    \lambda_2^o(s) & = & \bar\alpha_s
    \left(\Psi_{s'}(x') + \Psi_{s''}(x'') - \Psi_{s'}(x') \Psi_{s''}(x'') \right) +
    \bar\alpha_{s'}\Psi_{s'}(x') + \bar\alpha_{s''} \Psi_{s''}(x'').
  \end{IEEEeqnarray*}
  Let $z_{s}^o[0] = 0$, and for $i\geq 1$ let $z_{s}^o[i]$ be the maximum
  value of $z$ such that for any $x\in[0,\ z]$, we have
  \begin{IEEEeqnarray*}{rCl}
    F_{s}^o(x,z_{s'}^o[i-1],z_{s''}^o[i-1]) \geq 0,
  \end{IEEEeqnarray*}
  where $s\neq s' \neq s''$ and $s' < s''$.  The sequence
  $\{z_{s}^o[i]\}$ is increasing and upper bounded.  Let $z_{s}^{\circledast}$ be
  the limit of the sequence $\{z_{s}^o[i]\}$.
  Then %
  with probability converging to one, as $N\rightarrow \infty$, the
  ordinary BP decoding algorithm stops with at least $z_{s}^{\circledast} K_s$
  $s$-input packets being decoded for all $s\in \Theta$.
\end{theorem}
\begin{IEEEproof}
  The theorem will be proved as a special case of Theorem~\ref{the:L}.
\end{IEEEproof}

The following theorem tells us how many input packets are recovered
for each source node when the \emph{batched} BP decoding stops for binary LC-3
fountain codes.  

\begin{theorem}
  \label{the:l3}
  For each $s\in \Theta = \{\va,\vb,\vc\}$, fix $C_s>R_s>0$ and 
  consider a sequence of binary LC-$3$ fountain codes described above with $K_s/N
  \leq R_s$, $s\in \Theta$, $N=1,2,\ldots$. 
  For $s\neq s' \neq s''\in \Theta$, define
  \begin{IEEEeqnarray*}{rCl}
    F_s(x,x',x'')
    & = & \Psi_s'(x) + \frac{C_s/\beta_3}{\alpha_{s} + \lambda_1(s) + \lambda_2(s)}\ln(1-x),
  \end{IEEEeqnarray*}
  where
  \begin{IEEEeqnarray*}{rCl}
    \lambda_1(s) & = & 
    \alpha_{s+s'} \Psi_{s'}(x') + \alpha_{s+s''}\Psi_{s''}(x'') +
    \alpha_{s+s'+s''}\Psi_{s'}(x') \Psi_{s''}(x''),\\
    \lambda_2(s) & = & \bar\alpha \left(\Psi_{s'}(x') +
      \Psi_{s''}(x'') - \Psi_{s'}(x') \Psi_{s''}(x'') \right).
  \end{IEEEeqnarray*}
  Let $z_{s}[0] = 0$ and for $i\geq 1$ let $z_{s}[i]$ be the maximum
  value of $z$ such that for any $x\in[0,\ z]$, we have
  \begin{IEEEeqnarray*}{rCl}
    F_{s}(x,z_{s'}[i-1],z_{s''}[i-1]) \geq 0,
  \end{IEEEeqnarray*}
  where $s\neq s' \neq s''$ and $s' < s''$.  The sequence
  $\{z_{s}[i]\}$ is increasing and upper bounded.  Let $z_{s}^*$ be
  the limit of the sequence $\{z_{s}[i]\}$.
  Then %
  with probability converging to one, as $N\rightarrow \infty$, the
  batched BP decoding algorithm stops with at least $z_{s}^*K_s$
  $s$-input packets being decoded for all $s\in \Theta$.
\end{theorem}

\begin{remark}
  The performance of the batched BP decoding characterized in the
  above theorem does not depend on the individual values of 
  $\bar\alpha_{\va},\bar\alpha_{\vb},\bar\alpha_{\vc}$ as long as
  their summation is the same. 
\end{remark}

\begin{remark}
  In the above two theorems, $\lambda_2^o(s) \leq \lambda_2(s)$ for
  all $s$ and the inequalities are strict for at least 2
  users. Therefore, in general, $z_s^{\circledast} \leq z_s^*$ for all
  $s$ and the inequalities are strict for at least two users.
\end{remark}

\begin{IEEEproof}[Sketch of the proof]
  The theorem will be proved as a special case of Theorem~\ref{the:L}
  to be presented later.  Here we give a sketch of the proof. Compared
  with Theorem~\ref{the:1}, the major difference is the denominator of
  the second term of $F_s$. So we focus on how the denominator is
  obtained in this sketch.  The first stage of the batched BP decoding
  is similar to that of binary LC-2 fountain codes so we consider the
  second stage of the first round in the following.  Compared with the
  LC-2 fountain codes, we have more types of couples packets and
  non-autonomous output packets for LC-3 fountain codes.

  Consider an output packet $u = v_{\va} + v_{\vb} + v_{\vc}$, where
  $v_{s}$ is an $s$-coded packet. Packet $v_{\va}$ can be recovered as
  long as both $v_\vb$ and $v_\vc$ are linear combinations of the
  decoded input packets at the first stage.  So at the second
  stage of the first round, the probability that $v_{\va}$ can be
  recovered is at least $\Psi_{\vb}(z_{\vb}[1]) \Psi_{\vc}(z_{\vc}[1])$.

  Consider a batch formed by transfer matrix $H_{15}$ and coded
  packets $v_{\va}, v_{\vb}$ and $v_{\vc}$. If either $v_{\vb}$ or
  $v_{\vc}$ is a linear combination of the decoded input packets at
  the first stage, $v_{\va}$ can be recovered and used in the BP
  decoding in the next round. So at the second stage of the first
  round, the probability that $v_{\va}$ can be recovered by solving
  \eqref{eq:8} is at least $1-
  (1-\Psi_\vb(z_{\vb}[1]))(1-\Psi_\vc(z_{\vc}[1]))$.

  Counting all coupled
  $S$-output packets with $\va\in S$ and all the batches with
  transfer matrices $H_{15}$, $H_{16}$ and $H_{17}$, we get 
  that the number of $\va$-output packets recovered is at least 
  $n[\alpha_\va + \lambda_1(\va) + \lambda_2(\va)]$ at the second
  stage of the first round.
\end{IEEEproof}

For $s\neq s' \neq s''\in
\Theta$ with $s'<s''$, $F_s$ defined
in Theorem~\ref{the:l3} can be rewritten as
\begin{equation*}
  F_s(x,x',x'';C_s) = \Psi_s'(x) + \frac{C_s/\beta_3}{\Sigma(\Psi_{s'}(x'),\Psi_{s''}(x''))} \ln(1-x),
\end{equation*}
where
\begin{IEEEeqnarray*}{rCl}
  \Sigma(y,z) & = & \alpha_s + \alpha_{s+s'}
    y + \alpha_{s+s''} z + \alpha_{\Theta}yz +
    \bar{\alpha}\left(y+z-yz\right).
\end{IEEEeqnarray*}
Fixing one of the variables, $\Sigma(y,z)$ is an increasing function
of the other variable.  For $s\in \Theta$, define
\begin{equation*}
  f_{s}(x',x'') = f_s(x',x'';C_s) = \max \left\{z: F_{s}(x,x',x'') \geq 0, \ \forall x\in [0,
    z] \right\}.
\end{equation*}
The three sequences
$\{z_s[i]\}$, $s\in\Theta$ in Theorem~\ref{the:l3} satisfy
\begin{IEEEeqnarray*}{rCl}
  z_\va[i] & = & f_\va(z_{\vb}[i-1],z_\vc[i-1]), \\
  z_\vb[i] & = & f_\vb(z_{\va}[i-1],z_\vc[i-1]), \\
  z_\vc[i] & = & f_\vc(z_{\va}[i-1],z_\vb[i-1]).
\end{IEEEeqnarray*}
For $s\in\Theta$, function $f_s(\cdot,\cdot)$ is an increasing
function for both of its input variables.  The following lemma can be
proved by the monotonic property of the functions $f_s$, $s\in\Theta$.

\begin{lemma}\label{lem:sd3}
  The limit $(z_{\va}^*,z_{\vb}^*, z_{\vc}^*)$ of the three sequences defined in
  Theorem~\ref{the:l3} is the \emph{first} intersection of the surfaces
  $x = f_{\va}(y,z)$, $y = f_{\vb}(x,z)$ and $z = f_\vc(x,y)$, $x,y,z\in [0,1]$.
\end{lemma}
\begin{IEEEproof}
    This lemma is a special case of Lemma~\ref{lemma:inters}
  in Section~\ref{sec:LC}. 
\end{IEEEproof}

The definition of feasible points can be extended to LC-3 fountain
codes. For a point $(a_\va,a_\vb,a_\vc)$ in the region
$\{(x_\va,x_\vb,x_\vc):0\leq x_\va, x_\vb,x_\vc\leq 1\}$, we say
$(a_\va,a_\vb,a_\vc)$ is \emph{$(C_\va,C_\vb,C_{\vc})$-feasible} for
an LC-3 fountain code if $a_\va \leq f_{\va}(a_\vb,a_\vc;C_\va)$,
$a_\vb \leq f_{\vb}(a_\va, a_\vc;C_\vb)$ and $a_\vc \leq
f_{\vc}(a_\va,a_\vb;C_\vc)$.  The following theorem is useful in
deriving the degree-distribution optimization problems for binary LC-3
fountain codes.

\begin{theorem}
  \label{the:L23}
  For each $s\in\Theta=\{\va,\vb,\vc\}$, fix 
  $C_s>R_s>0$. Consider a sequence of binary LC-3
  fountain codes with $N=1,2\ldots$, where $K_s/N \leq
  R_s$ for $s\in\Theta$. For any $(a_\va,a_\vb,a_\vc)$, if there exists a
  feasible continuous curve $(x_\va(t),x_\vb(t),x_\vc(t))$ between the origin
  and
  $(a_\va,a_\vb,a_\vc)$, then i) a BP decoding algorithm will stop with at least
  $a_sK_s$ $s$-input packets being decoded for all $s\in\Theta$ with
  probability converging to one when $N\rightarrow \infty$, and ii)
  there exists an \emph{increasing} feasible continuous
  curve $(\tilde x_\va(t),\tilde x_\vb(t),\tilde x_\vc(t))$ between the origin and $(a_\va,a_\vb,a_\vc)$.
\end{theorem}
\begin{IEEEproof}
    This theorem is a special case of Theorem~\ref{the:L2} 
  in Section~\ref{sec:LC}. 
\end{IEEEproof}

\subsection{Optimizations}
\label{sec:degr-distr-optim}

Fix the parameters defined in Section~\ref{sec:batches}.  Suppose that we want to design a
binary LC-3 fountain codes such that at least $\eta_s$ fraction of
$s$-input packets can be decoded by the batched BP decoding for all
$s\in\Theta$. Theorem~\ref{the:L23} converts the problem to 
the existence of feasible curves: For any triple $\bar C =
(C_\va, C_\vb,C_\vc)$, if there exists a $\bar
C$-feasible curve between the origin and $(\eta_\va,\eta_\vb,
\eta_\vc)$, then the BP decoding will stop with at least $\eta_s K_s$
$s$-input packets decoded for all $s\in \Theta$, and hence the rate
triple $(\eta_\va C_\va, \eta_\vb C_\vb, \eta_\vc C_\vc)$ is \emph{achievable}
by the batched BP decoding.  Theorem~\ref{the:L23} also enables us to consider only
the increasing curves from the origin to $(\eta_\va,\eta_\vb,\eta_\vc)$.

By definition, a point $(\hat{x}_\va,\hat{x}_\vb,\hat{x}_\vc)$ is
$(C_\va,C_\vb,C_\vc)$-feasible if $\hat{x}_\va \leq
f_\va(\hat{x}_\vb,\hat{x}_\vc;C_\va)$, $\hat{x}_\vb \leq
f_\vb(\hat{x}_\va,\hat{x}_\vc;C_\vb)$ and $\hat{x}_\vc\leq
f_\vc(\hat{x}_\va,\hat{x}_\vb;C_\vc)$, which are equivalent to
\begin{IEEEeqnarray*}{rCl}
  F_\va(x,\hat{x}_\vb,\hat{x}_\vc;C_\va) & \geq & 0, \quad \forall x \in [0,\hat{x}_\va], \\
  F_\vb(x,\hat{x}_\va,\hat{x}_\vc;C_\vb) & \geq & 0, \quad \forall x \in [0,\hat{x}_\vb], \\
  F_\vc(x,\hat{x}_\va,\hat{x}_\vb;C_\vc) & \geq & 0, \quad \forall x \in [0,\hat{x}_\vc], 
\end{IEEEeqnarray*}
and hence equivalent to
\begin{IEEEeqnarray*}{rCl}
  \Sigma(\Psi_\vb(\hat{x}_\vb),\Psi_\vc(\hat{x}_\vc)) \Psi_\va'(x) +
  {C_\va/\beta_3} \ln(1-x) & \geq & 0, \quad \forall x \in
  [0,\hat{x}_\va], \\ 
  \Sigma(\Psi_{\va}(\hat{x}_\va),\Psi_\vc(\hat{x}_\vc))\Psi_\vb'(x) +
  {C_\vb/\beta_3} \ln(1-x) & \geq & 0, \quad \forall x \in
  [0,\hat{x}_\vb], \\
  \Sigma(\Psi_{\va}(\hat{x}_\va),\Psi_\vb(\hat{x}_\vb))\Psi_\vc'(x) +
  {C_\vc/\beta_3} \ln(1-x) & \geq & 0, \quad \forall x \in
  [0,\hat{x}_\vc].
\end{IEEEeqnarray*}
We only evaluate the zig-zag type of curves from the origin to
$(\eta_\va,\eta_\vb,\eta_\vc)$. Fix a positive integer $t_{\max}$ and
three sequences of real numbers 
$0=x_s[0]\leq x_s[1] \leq \ldots \leq x_s[t_{\max}] = \eta_s$, $s\in\Theta$. The curve
formed by line segments 
\begin{equation*}
  (x_\va[t],x_\vb[t],x_\vc[t]) - (x_\va[t+1],x_\vb[t],x_\vc[t]) -
  (x_\va[t+1],x_\vb[t+1],x_\vc[t]) - (x_\va[t+1],x_\vb[t+1],x_\vc[t+1])
\end{equation*}
$t=0, 1,\ldots, t_{\max}-1$ is an increasing zig-zag curve
from the origin to $(\eta_\va,\eta_\vb,\eta_\vc)$. Due to the property of the
feasible curves, we only need to check the feasibility of the points
\begin{equation}
  \label{eq:4}
\begin{IEEEeqnarraybox}[][c]{l}
  (x_\va[0],x_\vb[0],x_\vc[0]), (x_\va[1],x_\vb[0],x_\vc[0]),
  (x_\va[1],x_\vb[1],x_\vc[0]), \\ (x_\va[1],x_\vb[1],x_\vc[1]), 
  (x_\va[2],x_\vb[1],x_\vc[1]),\ldots,
  (x_\va[t_{\max}],x_\vb[t_{\max}],x_\vc[t_{\max}]).
\end{IEEEeqnarraybox}  
\end{equation}

We are now ready to introduce the optimization problems for binary
LC-3 fountain codes.
Fix $t_{\max}$, $C_\vb$, $C_\vc$, $\eta_{\va}$, $\eta_{\vb}$ and $\eta_\vc$.
The following optimization problem maximizes the achievable rate of
source node $\va$ for given rates of source nodes $\vb$ and $\vc$:
\begin{equation}
  \label{eq:op13}
    \begin{IEEEeqnarraybox}[][c]{r.l}
      \max & \eta_\va\theta_\va\beta_3 \\
      \text{s.t.} & 
       \forall s\in\Theta, x_s[0] = 0, x_s[t_{\max}] = \eta_{s};\\
       & \forall s\in\Theta, \forall t = 1, \ldots, t_{\max}, \quad
       x_s[t] \geq x_s[{t-1}]; \\
       & \forall t = 1, \ldots, t_{\max}, \\
       & \Sigma(\Psi_\vb({x}_\vb[t-1]),\Psi_\vc({x}_\vc[t-1])) \Psi_\va'(x) +
       {\theta_\va} \ln(1-x) \geq 0, \quad \forall x \in
       (x_\va[t-1],{x}_\va[t]], \\ 
       & \Sigma(\Psi_{\va}({x}_\va[t]),\Psi_\vc({x}_\vc[t-1]))\Psi_\vb'(x) +
       {C_\vb/\beta_3} \ln(1-x) \geq 0, \quad \forall x \in
       (x_\vb[t-1],{x}_\vb[t]], \\
       & \Sigma(\Psi_{\va}({x}_\va[t]),\Psi_\vb({x}_\vb[t]))\Psi_\vc'(x) +
       {C_\vc/\beta_3} \ln(1-x) \geq 0, \quad \forall x \in
       (x_\vc[t-1],{x}_\vc[t]],
    \end{IEEEeqnarraybox}
\end{equation}
where the variables of the optimization are $\theta_\va$, $x_s[t]$,
$t=1,\ldots,t_{\max}$, $s\in\Theta$, degree distributions $\Psi_\va$,
$\Psi_\vb$ and $\Psi_\vc$. The constraints of the above optimization
guarantee that the points in \eqref{eq:4} are feasible.

Fix $t_{\max}$, $\eta_{\va}$, $\eta_{\vb}$ and $\eta_\vc$.
The following optimization problem maximizes the sum rate of the three
source nodes:
\begin{equation}
  \label{eq:op23}
    \begin{IEEEeqnarraybox}[][c]{r.l}
      \max & \beta_3(\eta_\va\theta_\va+\eta_\vb\theta_\vb+\eta_\vc\theta_\vc) \\
      \text{s.t.} & 
       \forall s\in\Theta, x_s[0] = 0, x_s[t_{\max}] = \eta_{s};\\
       & \forall s\in\Theta, \forall t = 1, \ldots, t_{\max}, \quad
       x_s[t] \geq x_s[{t-1}]; \\
       & \forall t = 1, \ldots, t_{\max}, \\
       & \Sigma(\Psi_\vb({x}_\vb[t-1]),\Psi_\vc({x}_\vc[t-1])) \Psi_\va'(x) +
       {\theta_\va} \ln(1-x) \geq 0, \quad \forall x \in
       (x_\va[t-1],{x}_\va[t]], \\ 
       & \Sigma(\Psi_{\va}({x}_\va[t]),\Psi_\vc({x}_\vc[t-1]))\Psi_\vb'(x) +
       {\theta_\vb} \ln(1-x) \geq 0, \quad \forall x \in
       (x_\vb[t-1],{x}_\vb[t]], \\
       & \Sigma(\Psi_{\va}({x}_\va[t]),\Psi_\vb({x}_\vb[t]))\Psi_\vc'(x) +
       {\theta_\vc} \ln(1-x) \geq 0, \quad \forall x \in
       (x_\vc[t-1],{x}_\vc[t]],
    \end{IEEEeqnarraybox}
\end{equation}
where the variables of the optimization are $\theta_s$, $x_s[t]$,
$t=1,\ldots,t_{\max}$, $s\in\Theta$, degree distributions $\Psi_\va$, $\Psi_\vb$ and
$\Psi_\vc$.

\begin{remark}
The maximum degree $D$ can be similarly bounded as in Lemma~\ref{lem:4}.  
\end{remark}

\begin{remark}
We can similarly obtain the degree distribution optimization problems
for the ordinary BP decoding.
\end{remark}

\subsection{Achievable Rates}
\label{sec:numer-eval-1}

Given the distribution $g$ of the transfer matrix, we know from
Section~\ref{sec:performance-bounds} that a rate triple $(
R_\va, R_\vb, R_\vc)$ is achievable by the binary LC-3 fountain codes only if
\begin{IEEEeqnarray*}{rCl}
   R_\va & \leq & \sum_i g(H_i)\rank(H_i^{\{\va\}}) = g(H_1) + g(H_4) +
  g(H_5) + \sum_{i=7}^9g(H_i) + \sum_{i=11}^{15}g(H_i) \\
  & = &
  \beta_3(\alpha_\va+\alpha_{\va+\vb}+\alpha_{\va+\vc} +
  \alpha_{\va+\vb+\vc} + \bar\alpha), \\
   R_\vb & \leq & \sum_i g(H_i)\rank(H_i^{\{\vb\}}) = 
  \beta_3(\alpha_\vb+\alpha_{\va+\vb}+\alpha_{\vb+\vc} +
  \alpha_{\va+\vb+\vc} + \bar\alpha), \\
   R_\vc & \leq & \sum_i g(H_i)\rank(H_i^{\{\vc\}}) = 
  \beta_3(\alpha_\vc+\alpha_{\vb+\vc}+\alpha_{\va+\vc} +
  \alpha_{\va+\vb+\vc} + \bar\alpha), \\
   R_\va +  R_\vb & \leq & \sum_i g(H_i)
  \rank(H_i^{\{\va,\vb\}}) \\
  & = & g(H_1) + g(H_2) + 2 g(H_4) + g(H_5) + g(H_6) + 2 g(H_7) \\
  & & +  \sum_{i=8}^{11} g(H_i) + 2 g(H_{12}) + 2 g(H_{13}) + g(H_{14}) + 2
  g(H_{15}) \\
  & = & \beta_3(1-\alpha_\vc), \\
   R_\vb +  R_\vc & \leq & \sum_i g(H_i)
  \rank(H_i^{\{\vb,\vc\}})
   =  \beta_3(1-\alpha_\va), \\
   R_\va +  R_\vc & \leq & \sum_i g(H_i)
  \rank(H_i^{\{\va,\vc\}})
   =  \beta_3(1-\alpha_\vb), \\
   R_\va +  R_\vb +  R_\vc & \leq & \sum_i g(H_i)
  \rank(H_i) = \beta_3.
\end{IEEEeqnarray*}

Instead of specifying a value of $\beta_3$, we remove $\beta_3$ from
the objective functions of both \eqref{eq:op13} and \eqref{eq:op23} so
that the optimal values are the normalized (sum) rates. The
  best numerical results obtained by evaluating \eqref{eq:op23} are
  listed in Table~\ref{tab:3}, where we see that the normalized
  achievable sum rates are all close to $1$, the upper bound. One of
the vertex of the above region is
\begin{IEEEeqnarray*}{rCl}
 R_\va & = & \beta_3(
\alpha_\va+\alpha_{\va+\vb}+\alpha_{\va+\vc} + \alpha_{\va+\vb+\vc} +
\bar\alpha), \\
 R_\vb & = & \beta_3(\alpha_\vb+\alpha_{\vb+\vc}+\bar\alpha),\\
 R_\vc & = & \beta_3\alpha_\vc.
\end{IEEEeqnarray*}
We also evaluate \eqref{eq:op13} with $C_\vb/\beta_3 =
(\alpha_\vb+\alpha_{\vb+\vc}+\bar\alpha)/\eta_\vb$ and $C_\vc/\beta_3
= \alpha_\vc/\eta_\vc$.\footnote{We need to pick the parameters such
  that $(C_\vb,C_\vc)$ is an interior point of the projection of the
  capacity region on the plane $R_\va=0$.}  From
  Table~\ref{tab:3}, readers can verify that the normalized achievable
  rates of user $\va$ are all close to the corresponding values of
  $\alpha_\va+\alpha_{\va+\vb}+\alpha_{\va+\vc} + \alpha_{\va+\vb+\vc}
  + \bar\alpha$.

We also optimize the sum rate of the ordinary BP decoding and give the
best rates we obtained in Table~\ref{tab:3}. We see that the batched BP
decoding consistently achieves a sum rate above $95\%$ of the optimal
value, while the performance of the ordinary BP decoding decreases
significantly when $\bar \alpha$ becomes larger. For the normalized
rates given in Table~\ref{tab:3}, $\beta_3$ can be any value in
$(0,3)$.

\begin{table}
  \centering
  \caption{Achievable rates of binary LC-3 fountain codes for
    $\eta_\va = \eta_\vb = \eta_\vc = 0.98$.
    In both \eqref{eq:op13} and
    \eqref{eq:op23}, the objective functions are modified by removing
    $\beta_3$.
    $\hat R_\va/\beta_3$ is obtained
    by solving \eqref{eq:op13} with $C_\vb/\beta_3 =
    (\alpha_\vb+\alpha_{\vb+\vc}+\bar\alpha)/\eta_\vb$ and
    $C_\vc/\beta_3 = \alpha_\vc/\eta_\vc$, $\hat R_{\text{sum}}/\beta_3$ is obtained
    by solving \eqref{eq:op23}, and
    $\hat R_{\text{sum}}^o/\beta_3$ is obtained by solving a
    normalzied sum-rate
    maximization problem for the ordinary BP decoding.
    }
  \label{tab:3}
  \begin{tabular}{c|c|c|c|c|c|c}
    \hline
    $\alpha_\va,\alpha_\vb,\alpha_\vc$ &
    $\alpha_{\va+\vb},\alpha_{\va+\vc},\alpha_{\vb+\vc}$ & 
    $\alpha_{\va+\vb+\vc}$ &
    $\bar\alpha$ & $\hat R_\va/\beta_3$  & $\hat R_{\text{sum}}^o/\beta_3$ & $\hat
    R_{\text{sum}}/\beta_3$  \\
    \hline\hline
    0.2 & 0.1 & 0 & 0.05 & 0.4194 & 0.9592 & 0.9784 \\
    0.2 & 0 & 0.1 & 0.15 & 0.3957 & 0.9273 & 0.9775 \\
    0.1 & 0.1 & 0.1 & 0.15 & 0.4904 & 0.9099 & 0.9556 \\
    0.1 & 0 & 0.1 & 0.3 & 0.4521 & 0.8653 & 0.9636 \\
    0.05 & 0.05 & 0 & 0.35 & 0.4532 & 0.8466 & 0.9628 \\
    \hline
  \end{tabular}
\end{table}

\section{General LC Fountain Codes}
\label{sec:LC}

We now discuss general LC fountain codes for NCMA with $L$ users,
where the base field is not necessarily binary.  The coded
packets of a fountain code are not required to be generated
independently. Specifically, we relax the requirement that the degrees of the
coded packets are independent, and assume that the fraction of batches
with transfer matrix $H$ and the degree of the $s$-coded packet being
$d_s$ for all $s\in\Theta$ converges to
$g(H)\prod_{s\in\Theta}\Psi_s[d_s]$ as $N$ tends to infinity.

\subsection{Generalized Batched BP Decoders}

Both the ordinary BP decoder for LC-2 fountain codes and the batched
BP decoder for LC-3 fountain codes can be extended to decode LC-$L$ fountain
codes, $L > 3$. As discussed, both decoders can perform decoding
in rounds with each round having two stages. The first stage is the
same for both decoders, while the second stages are different. For
general LC-$L$ fountain codes, $L>3$, we have more options to process
the coupled output packets in the second stage. We first define a
generic (round-based batched BP) decoder of LC-$L$ fountain codes and
then discuss several instances of the generic decoder in terms of their
different operations in the second stage.

The generic decoder of LC-$L$ fountain codes
starts with the first round and each round has two stages:
\begin{itemize}
\item Stage 1: The ordinary BP decoding is applied on the $s$-output
  packets to decode the $s$-input packets. The decoding in the first
  stage is equivalent to the decoding of $L$ LT codes in
  parallel. The first stage of the first round uses the clean
    output packets decoded by the physical layer.
\item Stage 2: Each batch is processed individually by one of the
  algorithms to be specified later to recover a number of \emph{clean}
  output packets for the next round decoding. When no more clean
  output packets are recovered than the previous round, the decoding
  stops.
\end{itemize}

Now we discuss the instances of the generic decoder in terms of the
operations in the second stage, where the linear system of equations
in \eqref{eq:batch} is solved. In the following discussion,
  we fix $S \subset \Theta$ and assume that in \eqref{eq:batch}, the
  $r$-input packet $v_r$ has been decoded in the first stage if and
  only if $r\in S$.  We describe three instances of the generic
decoder.

The first instance of the generic decoder is the extension of the
ordinary BP decoder for LC-$2$ fountain codes, and is called the
\emph{BP-substitution decoder}.
The $i$-th row of $H$ is also called the $s$-th row where $s$ is the
$i$-th symbol in $\Theta$.  Denote by $H^S$ the
submatrix formed by the rows of $H$ indexed by $S$. The second stage
of the instance only substitutes the values of $v_r, r\in S$ into
\eqref{eq:batch} and obtain
\begin{equation}
  \label{eq:14}
  [v_s,s\in\Theta\setminus S] H^{\Theta\setminus S} = [u_1,\ldots,u_B] - [v_r, r\in S] H^S,
\end{equation}
where the LHS term is known.  Since no further operations are applied
to process the above linear system, for certain $s\in\Theta\setminus
S$, $v_s$ can be recovered if and only if $H^{\Theta\setminus S}$ has
a column where all the components are zero except for the component at
the $s$-th row. 

Both the second and third instances of the generic decoder can be
regarded as the extensions of the batched BP decoder for LC-3 fountain
codes.  The second instance is called the \emph{BP-BP decoder}, where
the (ordinary) BP algorithm is applied in the second stage.  The
operation in the second stage includes multiple iterations of the
following operations (see also Section~\ref{sec:bp}). The first
iteration is the same as the algorithm in the second stage of the
BP-substitution decoder. For each of the following iterations, the
clean output packets recovered in the last iteration are substituted
back into \eqref{eq:14} and new clear output packets are found (by
searching columns of $H^{\Theta\setminus S}$ with only one non-zero
component). Take \eqref{eq:8} as an example. Suppose that $v_\va$ is
known. The first iteration of the second stage will recover $v_\vc$ and
the second iteration of the second stage will recover $v_\vb$.

The third instance is called the \emph{BP-GE decoder}, where Gaussian
(Gauss-Jordan) elimination is applied in the second
stage. Specifically, in the second stage of the BP-GE decoder, the
substitution in the second stage of the BP-substitution decoder is
applied first. Following the substitution, Gaussian elimination
transforms $H^{\Theta\setminus S}$ into the reduced column echelon
form $\tilde H$. We then find the clean output packets by searching
columns of $\tilde H$ with only one non-zero component. To further
reduce the complexity, we can first apply the BP algorithm as in the
second stage of the BP-BP decoder and after the BP algorithm stops,
apply the Gaussian elimination.  Consider the following batch with
four users:
\begin{equation*}
  [u_1, u_2]
  =
  [v_\va, v_\vb, v_\vc, v_{\text{D}}]
  \begin{bmatrix}
    1 & 0 \\ 0 & 1 \\ 1 & 1 \\ 1 & 1
  \end{bmatrix}
\end{equation*}
where $v_\va, \ldots, v_{\text{D}}$ are the input packets. Suppose
that $v_\va$ is known. The second stage of the BP-BP decoder will stop
after the first iteration without any clean output packets
recovered. However, the second stage of the BP-GE decoder can recover
$v_\vb$.

 For the binary LC-2 fountain codes, the BP-substitution,
  BP-BP and BP-GE decoders are all the same as the ordinary BP decoder
  discussed in Section~\ref{sec:l2}. For the binary LC-3 fountain
  codes, the BP-substitution decoder is the same as the ordinary BP
  decoder discussed in Section~\ref{sec:l3}, and both the BP-BP and
  BP-substitution decoders are the same as the batched BP decoder
  discussed in Section~\ref{sec:l3}.

We evaluate the computation complexity of the BP-GE decoder of LC-$L$
fountain codes. The other two instances we discussed have lower
complexity. For a batch of
$r$ output packets, the complexity of Gaussian elimination for
recovering $r$ clean output packets is $O(r^3+rLT)$ finite-field
operations per batch. The total complexity to process all the batches
converges to
\begin{IEEEeqnarray*}{rCl}
  & & O\left(N\sum_H g(H) [\rank(H)^3+\rank(H)LT]\right)  \IEEEyesnumber \label{eq:cmx}\\
  & = &
  O\left(N (\beta_{L}L^2+\beta_{L}LT)\right) \\
  & = & \IEEEyesnumber \label{eq:cmx2}
  O\left(\bar n (L^2+LT)\right),
\end{IEEEeqnarray*}
where $\beta_{L}$ is defined in \eqref{eq:13} and $\bar n = N
\beta_{L}$ is the expected number of output packets.
The clean $s$-coded packets will be used in the BP decoding of
$s$-input packets, which has complexity $O(K_sT)$ finite-field
operations. Since $\bar n \geq \sum_s K_s$, the total decoding complexity
is dominated by \eqref{eq:cmx2}.

If we know that at most $\tilde L$ linear equations can be recovered by NCMA,
i.e., $\rank(H)\leq \tilde L$, the complexity \eqref{eq:cmx} can be
simplified to $O(\bar n (\tilde L^2+LT))$.

\subsection{Local Information Function}
\label{sec:gener-batch-bp}

Instead of analyzing the batched BP decoders defined above
individually, we provide a unified analysis of these decoders using
the following characterization of different algorithms in the second
stage.

Denote by
$\Theta^{\setminus s}$ the set $\Theta\setminus \{s\}$.
For a set $S$, denote by $2^S$ the collection of all subsets of $S$.
Recall that $\mathcal{H}_L$ is the collection of all the full-column-rank,
$L$-row matrices over $\ff_q$ (see Section~\ref{sec:performance-bounds}). 
For any
$s\in \Theta$, the \emph{local information function (LIF)}
$\gamma_s^*: \mathcal{H}_L \rightarrow 2^{\Theta^{\setminus s}}$ is defined by
\begin{enumerate}
\item for any $S \in \gamma_s^*(H)$, $v_s$ can be uniquely solved by
  \eqref{eq:batch} if the values of $v_r$, $r\in S$ are all known;
\item $\gamma_s^*(H)$ includes all such subsets of $\Theta^{\setminus s}$.
\end{enumerate}
In other words, for any
$S \in \gamma_s^*(H)$, using linear combinations of the equations in
\eqref{eq:batch}, we can obtain the equation
\begin{equation}\label{eq:10}
  v_s = u - \sum_{r\in S} \phi_r v_r,
\end{equation}
where $u$ is a linear combination of
$u_1,\ldots,u_{B}$, and $\phi_r \in \ff_q$.

Let us illustrate the definition of LIFs by several examples.  First
consider two special cases.  When the row of $H$ corresponding to
$v_s$ contains only `$0$'s, that is, $v_s$ is not involved in any
output packets of the batch, we have $\gamma_s^*(H)=\emptyset$. When
in one column of $H$, the component corresponding to $v_s$ is `1' and
the rest components are `0's, that is, one of the output packets in
the batch is exactly $v_s$, we have $\gamma_s^*(H) =
2^{\Theta^{\setminus s}}$, i.e., all the subsets of $\Theta^{\setminus
  s}$. Consider one more example with $\ff_q=\text{GF}(2)$,
$\Theta=\{\va,\vb,\vc,\mathrm{D}\}$, where $\va\leq \vb \leq \vc\leq
\mathrm{D}$, and the transfer matrix
\begin{equation}\label{eq:9}
  H =
  \begin{bmatrix}
    1 & 0 & 0 \\ 0 & 1 & 0 \\ 0 & 0 & 1 \\ 1 & 1 & 0
  \end{bmatrix}.
\end{equation}
We can see that
\begin{IEEEeqnarray*}{rCl}
  \gamma_\va^*(H) & = & 2^{\{\vb,\vc,\mathrm{D}\}}\setminus \{\{\vc\},\emptyset\},\\
  \gamma_\vb^*(H) & = & 2^{\{\va,\vc,\mathrm{D}\}}\setminus \{\{\vc\},\emptyset\},\\
  \gamma_\vc^*(H) & = & 2^{\{\va,\vb,\mathrm{D}\}},\\
  \gamma_{\mathrm{D}}^*(H) & = & 2^{\{\va,\vb,\vc\}}\setminus \{\{\vc\},\emptyset\}.
\end{IEEEeqnarray*}
We have the following basic properties of $\gamma_s^*$. 
\begin{lemma}\label{lemma:gamma}
  Let $H$ be an $L\times B$ full-column rank matrix over $\mathbb{F}_q$.
  \begin{enumerate}
\item If $S'\in \gamma_s^*(H)$, then $S\in
\gamma_s^*(H)$ for any $S'\subset S \subset
\Theta^{\setminus s}$;
\item $\gamma_s^*(H) = \gamma_s^*(H\Phi)$ for any full-rank $B\times
  B$ matrix $\Phi$.
\end{enumerate}
\end{lemma}

LIFs completely characterize the relations between $s$-coded packet
and other coded packets in a batch: The $s$-coded packet in a batch
with transfer matrix $H$ can be recovered by Gaussian elimination if
and only if for certain $S\in \gamma_s^*(H)$, all the values of $v_r,
r\in S$ are known.  We can also use certain subsets of $\gamma_s^*(H)$
to characterize the second stages of the BP-substitution and BP-BP
decoders.

A function $\gamma_s:
\mathcal{H}_L \rightarrow 2^{\Theta^{\setminus s}}$ is called a \emph{partial LIF} if
\begin{enumerate}
\item $\gamma_s(H) \subset \gamma_s^*(H)$;
\item for any $S\in \gamma_s(H)$, all
the super sets of $S$ in $\Theta^{\setminus s}$ are in $\gamma_s(H)$.
\end{enumerate}
For a subset $\mathcal{A}$ of $2^{\Theta^{\setminus s}}$, the span of
$\mathcal{A}$ in $\Theta^{\setminus s}$, denoted by
$\lspan{\mathcal{A}}_{\Theta^{\setminus s}}$, is the collection of all
$S\subset \Theta^{\setminus s}$ that include at least one element of
$\mathcal{A}$ as a subset.

Let us see an example of partial LIFs. For an $L\times B$ full-column rank
transfer matrix $H$, define $\text{supp}_j(H)$ for $1\leq j \leq B$ as
the support set of the $j$-th column of $H$, i.e., the subset of 
$s\in \Theta$ such that the component of $H$ on the $s$-th row, $j$-th column is
nonzero.  For the $H$ in \eqref{eq:9}, we have
\begin{IEEEeqnarray*}{rCl}
  \text{supp}_1(H) & = & \{\va,\mathrm{D}\}, \\
  \text{supp}_2(H) & = & \{\vb,\mathrm{D}\}, \\
  \text{supp}_3(H) & = & \{\vc\}.
\end{IEEEeqnarray*}
Define
\begin{equation*}
  \gamma_s^o(H) = \lspan{\{
  \text{supp}_i(H) \setminus \{s\}, i\in \{1,\ldots, B\}: s\in \text{supp}_i(H)\}}_{\Theta^{\setminus s}}.
\end{equation*}
We see that $\gamma_s^o$ is a partial LIF
since if $s\in \text{supp}_i(H)$ then $\text{supp}_i(H) \setminus
\{s\} \in \gamma_s^*(H)$. 
For the $H$ in
\eqref{eq:9}, we have
\begin{IEEEeqnarray*}{rCl}
 \gamma_\va^o(H) & = & \lspan{\{\{D\}\}}_{\{\vb,\vc,\text{D}\}}, \\ 
 \gamma_\vb^o(H) & = & \lspan{\{\{D\}\}}_{\{\va,\vc,\text{D}\}}, \\
 \gamma_\vc^o(H) & = & \lspan{\emptyset}_{\{\va,\vb,\mathrm{D}\}}, \\
 \gamma_{\mathrm{D}}^o(H) & = & \lspan{\{\{\va\},\{\vb\}\}}_{\{\va,\vb,\vc\}}.
\end{IEEEeqnarray*}
For a given linear system \eqref{eq:14} and $s\in\Theta$,
$\gamma_s^o(H)$ gives all the possible ways to solve $v_s$ without any
matrix operations. Therefore, $\gamma_s^o$ characterizes the second
stage of the BP-substitution decoder.

Let us continue to discuss how to characterize the second stage of the
BP-BP decoder. From the above discussion, $\gamma_s^o$ tells us the
solvability of $v_s$ using one iteration of the BP algorithm on
\eqref{eq:14}.  Define $\gamma_s^{b,1} = \gamma_s^o$.  For
$i=2,\ldots,L$, define function $\gamma_s^{b,i}:\mathcal{H}_L
\rightarrow 2^{\Theta^{\setminus s}}$ as
\begin{equation*}
  \gamma_s^{b,i}(H) = \gamma_s^{b,i-1}(H) \cup \left(\bigcup_{T\in
    \gamma_{s}^{o}(H)} \tilde\gamma_s^{i-1}(T,H)\right),
\end{equation*}
where
\begin{equation*}
  \tilde\gamma_s^{i-1}(T,H) = \left\{\bigcup_{r\in T}T_r: s\notin T_r\in
  \gamma_r^{b,i-1}(H),  \forall r\in T\right\}.
\end{equation*}
The following lemma tells that $\{\gamma_s^{b,i}, s\in \Theta\}$
characterizes the first $i$ iterations of the second stage of the
BP-BP decoder.

\begin{lemma}
  For $i=1,\ldots, L$, $\gamma_s^{b,i}$ are partial LIFs; and
  for $S\in \gamma_s^{b,i}(H)$, $v_s$ can be solved in terms of $v_r,
  r\in S$ using at most $i$ iterations of the ordinary BP algorithm on
  the linear system \eqref{eq:batch}. 
\end{lemma}
\begin{IEEEproof}
We prove the lemma by induction.
First the above claims hold for $i=1$. Fix $i>1$. For any $S\in
\gamma_s^{b,i}(H)$, either $S\in \gamma_s^{b,i-1}(H)$ or $S \in
\tilde\gamma_s^{i-1}(T,H)$ for certain $T\in \gamma_s^{o}(H)$.  If
$S\in \gamma_s^{b,i-1}(H)$, by the induction hypothesis, $v_s$ can be
solved using at most $i-1$ iterations of the ordinary BP algorithm,
and all the supersets of $S$ in $\Theta^{\setminus s}$ are in
$\gamma_s^{b,i-1}(H)$ and hence in $\gamma_s^{b,i}(H)$.  If $S \in
\tilde\gamma_s^{i-1}(T,H)$, then $S = \cup_{r\in T}T_r$ for certain
$T_r\in \gamma_r^{b,i-1}(H), s\notin T_r$. By induction hypothesis,
$v_r$ can be solved using at most $i-1$ iterations of the ordinary BP
algorithm in terms of $v_{r'}, r'\in T_r$. Since $T\in
\gamma_s^{o}(H)$, we can use one more iteration of the BP algorithm to
recover $v_s$ in terms of $v_{r'}, r'\in \cup_{r\in T} T_r$. Further,
for any $S'\supset S, S'\subset \Theta^{\setminus s}$, we can write
$S' = \cup_{r\in T}T_r'$, where $T_r' \supset T_r, r\in T$. Since
$s\notin T_r' \in \gamma_r^{b,i-1}(H)$, we have $S' \in
\tilde\gamma_s^{i-1}(T,H)$. This completes the proof of the lemma.
\end{IEEEproof}

We say a batched BP decoder of LC-$L$ fountain codes is characterized
by partial LIFs $\{\gamma_s, s\in\Theta\}$ if the second stage of each
round of the BP decoder satisfies the following property: For each
batch with transfer matrix $H$, known values of $v_r, r\in S$ and any
$s\in \Theta\setminus S$, $v_s$ can be recovered if and only if $S \in
\gamma_s(H)$. Specifically, the BP-substitution decoder, the BP-BP
decoder with $i$ iterations in the second stage, and the BP-GE decoder
are the bathed BP decoders characterized by $\{\gamma_s^o ,
s\in\Theta\}$, $\{\gamma_s^{b,i}, s\in \Theta\}$, and $\{\gamma_s^*,
s\in \Theta\}$, respectively. We will analyze a general batched BP
decoder characterized by any partial LIFs $\{\gamma_s, s\in\Theta\}$.

\subsection{Analysis of Decoding}

We analyze the performance of the batched BP decoder characterized by
partial LIFs $\{\gamma_s, s\in\Theta\}$. For $s\in \Theta$, transfer
matrix $H$ and $0\leq y_r \leq 1$, $r\in \Theta^{\setminus s}$, define
\begin{equation}\label{eq:7}
  \Gamma_s(H,y_{r},r\in \Theta^{\setminus s}) = \sum_{S\in
    \gamma_s(H)} \prod_{r\in S} y_r \prod_{r \in \Theta \setminus (\{s\}\cup S)}(1-y_{r}).
\end{equation}
Suppose that a batch is generated by $\{v_s,s\in\Theta\}$.  If with
probability $p_r$, the value of $v_r$ is known, then the probability
that $v_s$ can be expressed as the already-known $v_r, r\in
\Theta^{\setminus s}$ by the relations given in $\gamma_s(H)$ is
exactly $\Gamma_s(H,p_{r},r\in \Theta^{\setminus s})$. For example,
when $\gamma_s(H)=\emptyset$, the value of $\Gamma_s(H,p_{r},r\in
\Theta^{\setminus s})$ is zero; when $\gamma_s(H) =
2^{\Theta^{\setminus s}}$, the value of $\Gamma_s(H,p_{r},r\in
\Theta^{\setminus s})$ is one.

\begin{theorem}\label{the:L}
  For each $s\in\Theta$, fix $C_s> R_s > 0$.  Consider an LC-$L$
  fountain codes with $N$ batches employing a batched BP decoder
  characterized by partial LIFs $\{\gamma_s, s\in \Theta\}$, where
  $K_s/N \leq R_s$ for $s\in\Theta$. Define for $s\in \Theta$
  \begin{equation*}
    F_s(x,y_{r},r\in \Theta^{\setminus s}) = F_s(x,y_{r},r\in
    \Theta^{\setminus s};  C_s) = \Psi_s'(x) +
    \frac{ C_s}{\sum_{H}g(H)\Gamma_s(H,\Psi_r(y_{r}),r\in \Theta^{\setminus s})}\ln(1-x).
  \end{equation*}
  Let $z_{s}[0] = 0$ and for $i\geq 1$ let $z_{s}[i]$ be the maximum
  value of $z$ such that for any $x\in[0,\ z]$, we have
  \begin{IEEEeqnarray*}{rCl}
    F_{s}(x,z_{r}[i-1],r\in \Theta^{\setminus s}) \geq 0.
  \end{IEEEeqnarray*}
  The sequence $\{z_{s}[i]\}$ is increasing and upper bounded.  Let
  $z_{s}^*$ be the limit of the sequence $\{z_{s}[i]\}$.  Then there
  exists a positive number $c$ such that when $N$ is sufficiently
  large, with probability at least $1-e^{-cN}$, the batched
  BP decoder stops with at least $z_{s}^*K_s$
  $s$-input packets being decoded for all $s\in \Theta$.
\end{theorem}

\begin{remark}
  Since $\gamma_s^o(H)\subseteq \gamma_s^*(H)$ for all $s\in \Theta$, the value of  
$\Gamma_s$ with respect to $\gamma_s^*(H)$ is larger than or equal to the value of $\Gamma_s$ with respect to $\gamma_s^o(H)$. Therefore, in general the performance of batched BP decoding is better than the performance of ordinary BP decoding.
\end{remark}

The proof of the above theorem is postponed to the next
subsection. Let us show how to apply the above theorem to the binary LC-2
and LC-3 fountain codes. The binary LC-2 fountain code has four
non-trivial transfer matrices (see \eqref{eq:11}). The batched
BP decoder reduces to the ordinary BP decoder, i.e.,
$\gamma_s^*(H_i) = \gamma_s^o(H_i), i=1,\ldots,4$. 
We can calculate that for $\gamma_s = \gamma_s^*$,
\begin{IEEEeqnarray*}{rCl}
  \sum_ig(H_i)\Gamma_{\va}(H_i,y_{\vb}) & = & g(H_1) +
  g(H_3) y_{\vb} + g(H_4), \\
  \sum_ig(H_i)\Gamma_{\vb}(H_i,y_{\va}) & = & 
  g(H_2) + g(H_3) y_{\va} + g(H_4).
\end{IEEEeqnarray*}
Recall that $\beta_2 = g(H_1)+g(H_2)+g(H_3)+2g(H_4)$.  The proof of
Theorem~\ref{the:1} is completed by substituting
$\alpha_{\va}=\frac{g(H_1)+g(H_4)}{\beta_2}$,
$\alpha_{\vb}=\frac{g(H_2)+g(H_4)}{\beta_2}$ and $\alpha_{\va+\vb} =
\frac{g(H_3) }{\beta_2}$ into Theorem~\ref{the:L}.

The binary LC-3 fountain code has 17 non-trivial transfer matrices
(see Section~\ref{sec:batches}). The batched BP decoder of the binary
LC-3 fountain code is characterized by $\{\gamma_s^*, s
\in \{\va,\vb,\vc\}\}$.  Recall the parameters defined in
Section~\ref{sec:batches}.  We can calculate that when
$\gamma_s=\gamma_s^*$,
\begin{IEEEeqnarray*}{rCl}
  \sum_{i=1}^{17}g(H_i)\Gamma_{\va}(H_i, y_{\vb}, y_{\vc})/ \beta_3 & = & \alpha_\va
  + \alpha_{\va+\vb} y_{\vb} + \alpha_{\va+\vc}
  y_{\vc} + \alpha_{\va+\vb+\vc} y_{\vb}y_\vc +
  \bar\alpha(y_\vb+y_\vc-y_\vb y_\vc),  \\
  \sum_{i=1}^{17}g(H_i)\Gamma_{\vb}(H_i, y_{\va}, y_{\vc})/ \beta_3 & = & \alpha_\vb
  + \alpha_{\va+\vb} y_{\va} + \alpha_{\vb+\vc}
  y_{\vc} + \alpha_{\va+\vb+\vc} y_{\va}y_\vc +
  \bar\alpha(y_\va+y_\vc-y_\va y_\vc),  \\
  \sum_{i=1}^{17}g(H_i)\Gamma_{\vc}(H_i, y_{\va}, y_{\vb})/ \beta_3 & = & \alpha_\vc
  + \alpha_{\va+\vc} y_{\va} + \alpha_{\vb+\vc}
  y_{\vb} + \alpha_{\va+\vb+\vc} y_{\va}y_\vb +
  \bar\alpha(y_\va+y_\vb-y_\va y_\vb).
\end{IEEEeqnarray*}
The proof of Theorem \ref{the:l3} is completed by substituting the
above three equalities into Theorem~\ref{the:L}.

We now apply Theorem~\ref{the:L} to the binary LC-3 fountain code with
the ordinary BP decoding, which is characterized by
$\{\gamma_s^o,s \in \{\va,\vb,\vc\}\}$. 
We can calculate
that when $\gamma_s=\gamma_s^o$,
\begin{IEEEeqnarray*}{rCl}
  \sum_{i=1}^{17}g(H_i)\Gamma_{\va}(H_i, y_{\vb}, y_{\vc})/\beta_3 & = & \alpha_\va
  + \alpha_{\va+\vb} y_{\vb} + \alpha_{\va+\vc}
  y_{\vc} + \alpha_{\va+\vb+\vc} y_{\vb}y_\vc +
  \bar\alpha_{\va} (y_\vb+y_\vc-y_\vb y_\vc) + \bar\alpha_{\vb}y_\vb + \bar\alpha_{\vc}y_\vc,  \\
  \sum_{i=1}^{17}g(H_i)\Gamma_{\vb}(H_i, y_{\va}, y_{\vc}) /\beta_3 & = & \alpha_\vb
  + \alpha_{\va+\vb} y_{\va} + \alpha_{\vb+\vc}
  y_{\vc} + \alpha_{\va+\vb+\vc} y_{\va}y_\vc +
  \bar\alpha_{\va} y_\va + \bar\alpha_{\vb}(y_\va+y_\vb-y_\va y_\vc) + \bar\alpha_{\vc}y_\vc,  \\
  \sum_{i=1}^{17}g(H_i)\Gamma_{\vc}(H_i, y_{\va}, y_{\vb}) /\beta_3 & = & \alpha_\vc
  + \alpha_{\va+\vc} y_{\va} + \alpha_{\vb+\vc}
  y_{\vb} + \alpha_{\va+\vb+\vc} y_{\va}y_\vb +
  \bar\alpha_{\va}y_\va + \bar\alpha_{\vb}y_\vb+ \bar\alpha_{\vc}(y_\va+y_\vb-y_\va y_\vb).
\end{IEEEeqnarray*}
The proof of Theorem \ref{the:l3o} is completed by substituting the
above three equalities into Theorem~\ref{the:L}.

\subsection{Proof of Theorem~\ref{the:L}}

The proof of Theorem~\ref{the:L} uses an existing result for LT
codes. The following proposition is implied by \cite{Raptormono} and
can be proved using the AND-OR tree approach \cite{luby98}.

\begin{proposition}\label{thm:lt}
Fix $0<R<C\leq 1$.  Consider an LT code with $K$
input packets and $n \geq K/R$ coded packets, where the
empirical degree distribution of the coded packets converges in
probability to a degree
distribution $\Psi$ with a fixed maximum degree. For any
$0 < \eta < 1$, if 
\begin{equation}\label{eq:5}
  \Psi'(x) + C \ln(1-x) \geq 0,  \forall x \in [0,\eta],
\end{equation}
then there exists a positive number $c$ such that when $n$ is
sufficiently large,
with probability at least $1-\exp(-cn)$, the BP
decoder is able to recover at least $\eta K$ input packets.
\end{proposition}

\begin{IEEEproof}[Proof of Theorem~\ref{the:L}]
  In the analysis, we introduce
  an extra criterion to stop the first stage of each round: If
  the first stage does not stop after $K_sz_s[i]$ $s$-input packets
  have been decoded, we force the first stage to stop. 
  For $s\in\Theta$, define random variable $K_{s}[i]$ as the total
  number of decoded $s$-input packets after the $i$th round.
  We always have $K_s[i] \leq
  K_sz_s[i]$. We prove by induction that for a sufficiently large $N$
  and $i=1,2,\ldots,$
  \begin{equation}\label{eq:1}
    \Pr\left\{ K_s[i] = K_s z_{s}[i], s\in\Theta \right\} 
    =  1 - O(i\exp(-cN)).
  \end{equation}
  For a batch transfer matrix $H$, let $\Omega_H$ be the set of all
  batches with transfer matrix $H$. Define
  \begin{IEEEeqnarray*}{rCl}
    \delta_0 & = & 1 - \max_s ( R_s/ C_s)^{1/(L+1)}.
  \end{IEEEeqnarray*}
  Henceforth in the proof, we assume
  that
  \begin{equation}\label{eq:e1}
    |\Omega_H| \geq N g(H)(1-\delta_0), \text{ for all transfer
      matrices } H
  \end{equation}
  holds. Since $|\Omega_H|/N$ converges to $g(H)$ for
  all transfer matrix $H$, this assumption holds for sufficiently
  large $N$.

  We first prove \eqref{eq:1} for $i=1$. 
  Consider the first strage of the first round. Define
  \begin{equation*}
    U_H^0(s) =
    \begin{cases}
      \Omega_H & \text{if } \emptyset \in \gamma_s(H), \\
      \emptyset & \text{otherwise}.
    \end{cases}
  \end{equation*}
  We know that when $\emptyset\in \gamma_s(H)$, all $s$-coded packets
  embedded in the batches in $\Omega_H$ can be recovered and hence
  can be used in the BP decoding at the first round. Let
  \begin{equation*}
    U^0(s) = \cup_H U_H^0(s)
  \end{equation*}
  be the batches that can be used in the BP decoding of
  the $s$-input packets at the first round.
  For $s\in \Theta$ such
  that $|U^0(s)|=0$, we have $K_s[1] = 0$. Since
  $\emptyset\notin \gamma_s(H)$ for all $H$ in this case, we have
  $\sum_H g(H) \Gamma_s(H,0,\ldots,0) = 0$ and hence $z_s[1]=0$
  according to the definition in theorem. Therefore, $K_s[1] = K_s
  z_s[1]$.  Fix an $s\in \Theta$ such that $|U^0(s)|>0$.
  Since the empirical degree distribution of the $s$-coded packets
  embedded in batches in $U_H^0(s)$ converges to $\Psi_s$ when
  $U_H^0(s)\neq \emptyset$, we can apply Proposition~\ref{thm:lt} on
  the ordinary BP decoding of the $s$-coded packets embedded in the
  batches in $U^0(s)$. By \eqref{eq:e1}, we have
  \begin{equation*}
    |U^0(s)| \geq N\sum_H g(H) \Gamma_s(H,0,\ldots,0)
      (1 - \delta_0),
  \end{equation*}
  which implies
  \begin{equation*}
    \frac{K_s}{|U^0(s)|} \leq \frac{ R_s}{\sum_H g(H)
      \Gamma_s(H,0,\ldots,0) (1 - \delta_0)} <  \frac{ C_s}{\sum_H g(H)
      \Gamma_s(H,0,\ldots,0)}.
  \end{equation*}
  By the definition of $z_{s}[1]$ in the theorem, we see
  that \eqref{eq:5} holds with $z_{s}[1]$, $\Psi_s$ and $\frac{ C_s}{\sum_H g(H)
      \Gamma_s(H,0,\ldots,0)}$ in place of $\eta$, $\Psi$ and $C$,
    respectively, and
  hence \eqref{eq:1} with $i=1$ is proved by Proposition~\ref{thm:lt}
  and the union bound.

  Assume that \eqref{eq:1} holds for certain $i\geq 1$.
  Suppose that after the first stage of the $i$-th round,
  \begin{equation}
    \label{eq:6}
    K_{s}[i] = K_s z_{s}[i], \forall s\in\Theta,
  \end{equation}
  which holds with probability at least $1 - O(i\exp(-cN))$ by
  the induction hypothesis. Suppose that the set
  $U_H^{i-1}(s)$ has been assigned, and only the batches in
  $U^{i-1}(s):=\cup_H U_H^{i-1}(s)$ are used in the decoding of the
  $s$-input packets at the first stage of the $i$-th round.

  Consider the second stage of the $i$-th round. For a batch $b$,
  denote by $v_s(b)$ the $s$-coded packet embedded in the batch.  We
  say that $v_{s}(b)$ is \emph{BP decodable after $i$-th rounds} if
  $v_s(b)$ is the linear combination of the decoded $s$-input packets
  in the first stage of the $i$-th rounds.  Denote by $p_s[i]$ the probability that
  for a randomly selected batch $b\notin U^{i-1}(s)$, $v_s(b)$ is BP
  decodable after the $i$-th round of decoding.  Since the neighbors
  of a coded packets are chosen uniformly at random, conditioning on
  the event in \eqref{eq:6}, we have
  \begin{IEEEeqnarray*}{rCl}
    p_{s}[i] \geq \sum_{d}\Psi_{s}[d](1-\delta_0/2) \frac{\binom{K_s
        z_{s}[i]}{d}}{\binom{K_s}{d}} 
    & \geq & \IEEEyesnumber \label{eq:3}
    \Psi_{s}(z_{s}[i]) (1 - \delta_0),
  \end{IEEEeqnarray*}
  where the inequalities hold for sufficiently large $K_s$. 
  Let $\delta_1 = \min_s(1/\Psi_{s}(z_s^*)-1)\delta_0$.
  On the
  other hand, we have  for sufficiently large $K_s$,
  \begin{IEEEeqnarray*}{rCl}
    p_{s}[i] \leq \sum_{d}\Psi_{s}[d] (1+\delta_1)\frac{\binom{K_s
        z_{s}[i]}{d}}{\binom{K_s}{d}} 
    & \leq & 
    \Psi_{s}(z_{s}[i])(1+\delta_1),
  \end{IEEEeqnarray*}
  which implies
  \begin{equation} \label{eq:42i}
    1 - p_{s}[i] \geq 1 - \Psi_{s}(z_{s}[i])(1+\delta_1) \geq (1-\Psi_s(z_s[i]))(1-\delta_0).
  \end{equation}

  For a set $U$, let $\text{Sa}(U,p)$ be a subset of $U$ where each element
  in $U$ is chosen with probability $p$ independently.
  Define
  \begin{IEEEeqnarray*}{rCl}
    D_H^i(s) & = & \left\{b \in \Omega_H\setminus
       U_H^{i-1}(s): v_s(b)\text{ is BP decodable after $i$-th rounds}\right\}
     \cup \text{Sa}(U_H^{i-1}(s), p_s[i]),\\
    D_H^i(s,S) & = & \cap_{r\in S} D_H^i(r) \setminus \cup_{r'\notin
      \{s\}\cup S} D_H^i(r'), \quad S \subset \Theta^{\setminus s},\\
    U_H^i(s) & = & \cup_{S\in \gamma_s(H)} D_H^i(s,S).
  \end{IEEEeqnarray*}
  For a batch
  $b\in D_H^i(r)$, the $r$-coded packet embedded in $b$ is either BP
  decodable after $i$ rounds (when $b\notin U_H^{i-1}(r)$) or known
  before the $i$-th rounds (when $b\in U_H^{i-1}(r)$).  So for
  $s\notin S\subset \Theta$ and batch
  $b \in D_H^i(s,S)$, all $v_r(b), r\in S$ are known after the first
  stage of the $i$-th
  round. If we further have $S\in \gamma_s(H)$, $v_s(b)$ can be recovered
  in terms of $v_r$, $r\in S$. Therefore, for all the batches $b$ in $
  U_H^{i}(s)$, the $s$-coded packets embedded in $b$ can be recovered
  at the second stage of the $i$-th round,
  and hence can be used in the BP decoding of the $(i+1)$-th round.

  Turn to the first stage of the $(i+1)$-th round. Let $U^{i}(s) =
  \cup_H U_H^i(s)$.  To apply Proposition~\ref{thm:lt} on the ordinary
  BP decoding of the $s$-input packet at the $(i+1)$-th round, we need
  to verify the degree distribution of the $s$-coded packets recovered
  from the batches in $U^{i}(s)$ and count the cardinality of
  $U^{i}(s)$.  Each batch $b\in \Omega_H$ is in $D_H^i(s,S)$
  independently with probability $\prod_{r\in S}p_r[i] \prod_{r'\notin
    \{s\}\cup S}(1-p_{r'}[i])$. So the degree distribution of the
  $s$-coded packets embedded in the batches in $D_H^i(s,S)$ converges
  in probability to $\Psi_s$ as $N$ tends to infinity.  Since
  $D_H^i(s,S)$, $S\subset \Theta^{\setminus s}$ form a partition of
  $\Omega_H$, we have
  \begin{equation*}
    |U^{i}(s)| = \sum_{H}\sum_{S\in \gamma_s(H)} |D_H^i(s,S)|,
  \end{equation*}
  and hence
  \begin{equation*}
    \E[|U^{i}(s)|] = \sum_H |\Omega_H| \Gamma_s(H,p_r[i],r\in \Theta^{\setminus s}).
  \end{equation*}
  
    Define event $E_N^i$ as
  \begin{equation*}
    |U^{i}(s)| \geq \sum_H |\Omega_H| \Gamma_s(H,p_r[i],r\in
    \Theta^{\setminus s}) ( 1 - \delta_0), \ \forall s\in\Theta.
  \end{equation*}
  By the Chernoff bound, event $E_{N}^i$ holds with probability at
  least $1 - O(\exp(-c(\delta_0)N))$, where $c(\delta_0)>0$ is a function
  of $\delta_0$.  
  Under
  the condition that the event $E_{N}^i$ holds, together with \eqref{eq:e1},
  \eqref{eq:3} and \eqref{eq:42i}, we have
  \begin{equation*}
    |U^{i}(s)| \geq N\sum_H
    g(H)\Gamma_s(H,\Psi_r(z_r[i]),r\in\Theta^{\setminus s})(1-\delta_0)^{L+1},
  \end{equation*}
  which implies
  \begin{equation*}
    \frac{K_s}{|U^{i}(s)|} \leq \frac{ R_s}{\sum_H
    g(H)\Gamma_s(H,\Psi_r(z_r[i]),r\in\Theta^{\setminus
      s})(1-\delta_0)^{L+1}}
    < \frac{ C_s}{\sum_H
    g(H)\Gamma_s(H,\Psi_r(z_r[i]),r\in\Theta^{\setminus
      s})}.
  \end{equation*}
  By the definition of $z_s[i+1]$ in the theorem, we see that
  \eqref{eq:5} holds with $z_{s}[i+1]$, $\Psi_s$ and $\frac{
    C_s}{\sum_H g(H) \Gamma_s(H,\Psi_r(z_r[i]),r\in\Theta^{\setminus
      s})}$ in place of $\eta$, $\Psi$ and $C$, respectively.
  By Proposition~\ref{thm:lt}, when $N$ is sufficiently large, we have
  \begin{equation*}
    \Pr\left\{ K_s[i+1] \geq K z_s[i+1]| K_{r}[i] = K_r z_{r}[i], \forall r\in\Theta \right\} 
    =  1 - O(\exp(-cN)),
  \end{equation*}
  where the probability that event $E_{N}^i$ holds is counted by
  modifying $c$.  Using the union bound and counting the probability
  that \eqref{eq:6} holds, \eqref{eq:1} is proved with $i+1$ in place of
  $i$.

  We only need to run at most $\sum_sK_s$ rounds of the decoding
  algorithm before no new input packets can be decoded. Therefore,
  with probability $1 - O(N\exp(-cN))$,
  a BP decoding algorithm stops with at least $z_{s}K_s$ $s$-variable
  decoded for all $s\in \Theta$. The proof is completed by decreasing
  $c$ slightly. 
\end{IEEEproof}

\subsection{Geometric Characterization}

For $s\in \Theta$, define
\begin{equation*}
  f_{s}(y_r, r\in \Theta^{\setminus s}) = f_{s}(y_r, r\in \Theta^{\setminus s};  C_s) = \max \left\{z:
    F_{s}(x,y_r, r\in \Theta^{\setminus s}; C_s) \geq 0, \ \forall x\in [0,
    z] \right\}.
\end{equation*}
The sequences
$\{z_s[i]\}$, $s\in\Theta$ defined in Theorem~\ref{the:L} satisfy
\begin{IEEEeqnarray*}{rCl}
  z_s[i] & = & f_s(z_{r}[i-1],r\in \Theta^{\setminus s}).
\end{IEEEeqnarray*}
With the help of the following lemma, we see that $f_s$ is an
increasing function for all the input variables. 

\begin{lemma}
  For any $t\in \Theta^{\setminus s}$, $\Gamma_s(H,p_r,r\in
  \Theta^{\setminus s})$ is an increasing function of $p_t$ with any given
  values of $p_r\in [0, 1]$, $r\in \Theta\setminus\{s,t\}$. 
\end{lemma}
\begin{IEEEproof}
  First, for all $S\in \gamma_s(H)$ with $t \in S$, the derivative of
  $\prod_{r\in S} p_r \prod_{r'\notin \{s\}\cup S}(1-p_{r'})$ for $p_t$ is nonnegative.
  Suppose that $S\in \gamma_s(H)$, $t\notin S$. Since $S\cup\{t\} \in
  \gamma_s(H)$, by definition $\Gamma_s(H,p_r,r\in \Theta^{\setminus s})$
  includes the summation of two terms:
  \begin{IEEEeqnarray*}{Cl}
    \prod_{r\in S} & p_r \prod_{r'\notin S\cup\{s\}}(1-p_{r'}), \\
    \prod_{r\in S\cup\{t\}} & p_r \prod_{r'\notin S\cup\{s,t\}}(1-p_{r'}).
  \end{IEEEeqnarray*}
  The derivatives of these two terms for $p_t$ are 
  \begin{IEEEeqnarray*}{rCl}
    - & \prod_{r\in S} & p_r \prod_{r'\notin S\cup\{s,t\}}(1-p_{r'}), \\
     & \prod_{r\in S} & p_r \prod_{r'\notin S\cup\{s,t\}}(1-p_{r'}),
  \end{IEEEeqnarray*}
  respectively.  Since the summation of these two derivatives is zero,
  the derivative of $\Gamma_s(H,p_r,r\in \Theta^{\setminus s})$ for $p_t$ is
  nonnegative.
\end{IEEEproof}

The following lemma gives a geometric characterization of the limits
of the sequences $\{z_s[i]\}$, $s\in\Theta$ defined in Theorem~\ref{the:L}.

\begin{lemma}\label{lemma:inters}
  The point $(z_{s}^*,s\in\Theta)$ of the limits of the sequences defined in
  Theorem~\ref{the:L} is an intersection of the surfaces
  $y_s = f_{s}(y_r,r\in\Theta^{\setminus s})$, $s\in \Theta$, and for
  any point $(x^*_r,r\in\Theta)$ on the intersection of 
  $y_s = f_{s}(y_r,r\in\Theta^{\setminus s})$, $s\in \Theta$,
  $z_s^*\leq x_s^*$ for all $s\in\Theta$. In other words,
  $(z_{s}^*,s\in\Theta)$  is   the \emph{first} intersection of the surfaces
  $y_s = f_{s}(y_r,r\in\Theta^{\setminus s})$, $s\in \Theta$.
\end{lemma}
\begin{IEEEproof}
  The lemma can be proved using the monotonic property of functions
  $f_s$.  Since $(z_{s'}[i], s'<s, z_{s}[i+1], z_{s''}[i], s''>s)$ is
  on $y_s=f_{s}(y_r,r\in\Theta^{\setminus s})$ for all $s\in\Theta$,
  the limit point $(z_{s}^*,s\in\Theta)$ is on $y_s =
  f_{s}(y_r,r\in\Theta^{\setminus s})$ for all $s\in \Theta$. The
  existence of the intersections of $y_s =
  f_{s}(y_r,r\in\Theta^{\setminus s})$ for all $s\in \Theta$ is
  guaranteed by the existence of the limits of the sequences
  $\{z_s[i]\}$, $s\in\Theta$.
  
  Let $(x^*_r,r\in\Theta)$ be an intersection of $y_s =
  f_{s}(y_r,r\in\Theta^{\setminus s})$ for all $s\in \Theta$. We show
  that $z_{s}[i]\leq x_s^*$, $s\in \Theta$ by induction. First, by
  definition $z_{s}[0] = 0 \leq x_s^*$, $s\in \Theta$. Assume that
  $z_{s}[j]\leq x_s^*$, $s\in\Theta$ for some $j\geq 0$. Since $f_{s}$
  is an increasing function of all the input variables, we have
  $z_{s}[j+1] = f_{s}(z_{r}[j],r\in\Theta^{\setminus s}) \leq
  f_{s}(x_r^*,r\in\Theta^{\setminus s}) = x_s^*$. Therefore, $z_s^*
  \leq x_s^*$ for all $s\in\Theta$, and hence the first intersection
  is well defined.
\end{IEEEproof}

Let ${\mathbf{C}} = ( C_s, s\in\Theta)$.  We say a point
$(a_r,r\in\Theta)$ in the region $\{(x_r,r\in\Theta):0\leq x_r \leq
1\}$ is \emph{${\mathbf{C}}$-feasible} for an LC-$L$ fountain
code if $a_s \leq f_{s}(a_r,r\in\Theta^{\setminus s}; C_s)$. 
A curve is \emph{${\mathbf{C}}$-feasible} for an LC-$L$ fountain
code if every point on the curve is ${\mathbf{C}}$-feasible.
A point/curve is said to
be \emph{feasible} when ${\mathbf{C}}$ is implied.
 One property
of feasible points is that if both $(a_r,r\in\Theta)$ and
$(b_r,r\in\Theta)$ are ${\mathbf{C}}$-feasible, where $a_s>b_s$
and $a_r=b_r$ for $r\in\Theta^{\setminus s}$, then the segment between
these two points is ${\mathbf{C}}$-feasible.  The reason is that
for any $x \in (b_s,a_s)$, we have $x \leq a_s \leq
f_{s}(a_r,r\in\Theta^{\setminus s})$ and for $r\in\Theta^{\setminus
  s}$, $a_r = b_r \leq f_{r}(b_{t},t\in \Theta^{\setminus r}) \leq
f_{r}(b_t,t\neq r < s, x, b_{t'}, t'\neq r > s)$ (since $f_{r}$ is an
increasing function for all input variables).

\begin{theorem}
  \label{the:L2}
  For each $s\in\Theta$, fix $C_s > R_s > 0$. Consider an LC-$L$
  fountain codes with $N$ batches employing a batched BP decoder
  characterized by
  partial LIFs $\{\gamma_s, s\in\Theta\}$, where $K_s/N \leq
  R_s$ for $s\in\Theta$. For any point $(a_r,r\in\Theta)$, if there
  exists a ${\mathbf{C}}$-feasible continuous curve
  $(x_r(t),r\in\Theta)$ between the origin and $(a_r,r\in\Theta)$,
  then i) the batched BP decoder will stop with
  at least $a_sK_s$ $s$-input packets decoded for all $s\in\Theta$
  with probability at least $1-e^{-cN}$ when $N$ is sufficiently
  large, where $c$ is a constant value, and ii) there exists an
  \emph{increasing} ${\mathbf{C}}$-feasible continuous curve $(\tilde
  x_r(t),r\in\Theta)$ between the origin and $(a_r,r\in\Theta)$.
\end{theorem}
\begin{IEEEproof}
  Suppose there exists a feasible continuous curve
  $V(t)=(x_r(t),r\in\Theta)$ between the origin and
  $(a_r,r\in\Theta)$.  We first prove ii) by constructing an
  increasing feasible continuous curve $(\tilde
  x_r(t),r\in\Theta)$. For a given $s\in\Theta$, we will show in the
  next paragraph that we can modify $V(t)$ to a feasible continuous
  curve $V_s(t)=(x_r'(t),r\in\Theta)$ between the origin and
  $(a_r,r\in\Theta)$ where $x_s'(t)$ is an increasing function of $t$
  and for $r\neq s$, $x_r'(t)=x_r(t)$. Then we can apply the above
  modification to all the coordinations successively to obtain an
  increasing feasible continuous curve $(\tilde x_r(t),r\in\Theta)$
  between the origin and $(a_r,r\in\Theta)$.
  
  Find the smallest $t'$ such that $x_s(t')=a_s$. We modify
  $(x_r(t),r\in\Theta)$ by replacing the part after $t=t'$ with a line
  segment between $(x_r(t'), r\in \Theta)$ and $(a_r,r\in\Theta)$
  without changing $x_r(t)$, $t \geq t'$ for all $r\neq s$.
  The new curve is still feasible and continuous and has the same
  parametric coordinate functions for all the positions other than
  $s$. We use the same notation for the coordination function at the
  position of $s$. The curve $V_s(t)$ is then formed as follows: start from
  $t=0$, $V_s(t)$ is the same as $V(t)$ until $t$
  increases to $\tau$ such that $x_s(\tau)$ is a local maximum point
  of $x_s(t)$. Find $\tau'$ as the largest $t \geq \tau$ such that
  $x(t)=x(\tau)$. We extend $V_s(t)$ from $(x_r(\tau),r\in \Theta)$ to
  $(x_r(\tau'),r\in\Theta)$ by a line segment without changing
  $x_r(t)$, $\tau \leq t \leq \tau'$ for all $r\neq s$. Repeat the above
  procedure from $t=\tau'$ until the end of the curve is
  reached.  We see that $x'_s(t)$ is increasing and ends
  at $a_s$, and for $r\neq s$, $x_r'(t)=x_r(t)$. This completes the
  proof of ii).

  We prove i) by assuming that curve $V(t)$ is
  increasing. Fix any $\tilde{C}'_r$ such that $\tilde{R}_r<\tilde{C}'_r<\tilde{C}_r$
  for all $r\in\Theta$. Let
  $(b_r,r\in\Theta)$ be any intersection of $y_s =
  f_{s}(y_r,r\in\Theta^{\setminus s}; \tilde{C}')$, $s\in \Theta$.
  If $b_r \geq a_r$ for all $r\in\Theta$, the claim of the theorem
  holds by Lemma~\ref{lemma:inters} and Theorem~\ref{the:L}.  In the
  following, we show by contradiction that it is not possible that
  $b_r < a_r$ for certain $r\in\Theta$.
  Without loss of generality, suppose that for certain $s\in\Theta$,
  $b_r<a_r$ for all $r\leq s$ and $b_r\geq a_r$ for all $r>s$. Since
  the curve $V(t)$ is increasing, continuous and ends
  at $(a_r,r\in\Theta)$, it must cross a point $(c_r,r\in\Theta)$
  satisfying
  \begin{enumerate}
  \item $c_{t} = b_{t} < a_{t}$ for certain $t\leq s$,
  \item $c_{r} \leq b_{r} < a_r$ for $r\neq t \leq s$, and 
  \item $c_r \leq a_r \leq b_r$ for all $r>s$.
  \end{enumerate}
  We have
  \begin{IEEEeqnarray}{rCl}
    c_{t} & = & \IEEEnonumber
    b_{t} \\
    & = & \label{eq:ic2}
    f_{t}(b_r,r\neq t \leq s, b_{r'},r'>s;{C}') \\
    & \geq & \label{eq:ic3}
    f_{t}(c_r,r\neq t \leq s, c_{r'},r'>s;{C}') \\
    & > & \label{eq:ic4}
    f_{t}(c_r,r\neq t \leq s, c_{r'},r'>s;{C}),
  \end{IEEEeqnarray}
  where \eqref{eq:ic2} follows that $(b_r,r\in\Theta)$ is on
  $y_t= f_t(\cdots; {C}')$; and \eqref{eq:ic3} and
  \eqref{eq:ic4} are obtained using
  the monotonic property of $f_t$. Since \eqref{eq:ic4} implies that
  $(c_r,r\in\Theta)$ is not feasible, we obtain a contradiction to
  that $(c_r,r\in\Theta)$ is on $V(t)$. Therefore, $a_r\leq b_r$ for
  all $r$ and the proof is completed.
\end{IEEEproof}

\section{Concluding Remarks}
\label{sec:conc}

Motivated by NCMA, we analyzed and designed near optimal
linearly-coupled fountain codes for linear multiple-access
channels. The coupling of codes is a general phenomenon when network
coding is used in a network with multiple source nodes.  To the best
of our knowledge, our work provides the first analysis of the joint BP
decoding of messages from multiple sources coupled by network
coding. Leveraging on the simplicity of batched BP decoding, our
framework may find application in many practical multi-source
communication systems besides NCMA.

\appendix[Solving the Optimization Problems]

The optimization problems \eqref{eq:op1}, \eqref{eq:op2},
\eqref{eq:op13} and \eqref{eq:op23} are in general non-convex.  We
take optimization problem \eqref{eq:op2} as an example to present how
to numerically solve these optimization problems. The variables of the
optimization are $\theta_\va$, $\theta_\vb$, $x_t$, $y_t$,
$t=1,\ldots,t_{\max}$, degree distributions $\Psi_\va$ and $\Psi_\vb$.
Consider the non-linear constraint
\begin{equation}\label{eq:conexample}
  \left(\alpha_{\va} + \alpha_{\va+\vb}
         \Psi_{\vb}\left(y_{t-1}\right)\right)\Psi_{\va}'(x) +
       \theta_{\va} \ln(1-x)\geq 0,\quad \forall x \in (x_{t-1}, x_t].
\end{equation}
Since it is impossible to check the inequality for all $x\in (x_{t-1},
x_t]$, we interpolate a number of $M$ (e.g., 20) points that are evenly
distributed in $(x_{t-1},x_t]$, and force them to satisfy the above
inequality.  The same relaxation is applied to other non-linear
constraints.  We then solve this (relaxed) optimization using a
non-linear optimization solver.\footnote{We use the \emph{fmincon}
  function provided in Matlab with the active-set algorithm.}

\begin{figure}
 \centering
 \begin{subfigure}[b]{0.4\textwidth}
    \centering
    \begin{tikzpicture}
      \begin{axis}[
        xlabel = $x$, ylabel= $y$,
        width=220, height=200,
        xmin=0,xmax=1, ymin=0,ymax=1,
        legend pos= north west, legend style={font=\small},
        label style={font=\small},
        mark size={1.0},
        grid=both]
        \addplot[color=blue, thick, no marks] table[x=max_x1, y=y_index] {fAB2.txt};
        \addplot[color=red, very thick, densely dotted, no marks]table[x=y_index, y=max_x2] {fAB2.txt};
        \addplot[color=black,mark=o] table[x=xA, y=xB] {xAB2.txt};
        \legend{$f_{\va}(y;\theta_\va)$, $f_{\vb}(x;\theta_\vb)$, zig-zag curve}
      \end{axis}
    \end{tikzpicture}
    \caption{Full figure}
  \end{subfigure}
  \qquad\qquad
   \begin{subfigure}[b]{0.4\textwidth}
    \centering
    \begin{tikzpicture}
      \begin{axis}[
        xlabel = $x$, ylabel= $y$,
        width=220, height=200,
        xmin=0,xmax=1, ymin=0,ymax=0.15,
        legend pos= north west, legend style={font=\small},
        label style={font=\small},
        mark size={1.0},
        grid=both]
        \addplot[color=blue, thick, no marks] table[x=max_x1, y=y_index] {fAB2.txt};
        \addplot[color=red, very thick, densely dotted, no marks]table[x=y_index, y=max_x2] {fAB2.txt};
        \addplot[color=black,mark=o] table[x=xA, y=xB] {xAB2.txt};
        \legend{$f_{\va}(y;\theta_\va)$, $f_{\vb}(x;\theta_\vb)$, zig-zag curve}
      \end{axis}
    \end{tikzpicture}
    \caption{Zoom in}
  \end{subfigure}
  \caption{Curves $x = f_{\va}(y;\theta_\va)$ and $y = f_{\vb}(x;\theta_\vb)$ with
   $\alpha_{\va}=\alpha_{\vb}=0.25$ and $\eta_\va = \eta_\vb = 0.98$,
   where $\theta_\va$, $\theta_\vb$, $x_t$, $y_t$,
$t=1,\ldots,t_{\max}$, $\Psi_\va$ and $\Psi_\vb$ are returned by the
optimization solver. The zig-zag curve formed by line segments $(x_t,y_t) - (x_{t+1},y_t) - (x_{t+1},
y_{t+1})$, $t=0, 1,\ldots, t_{\max}-1$. Here $M=10$. }\label{fig:infeasible}
\end{figure}
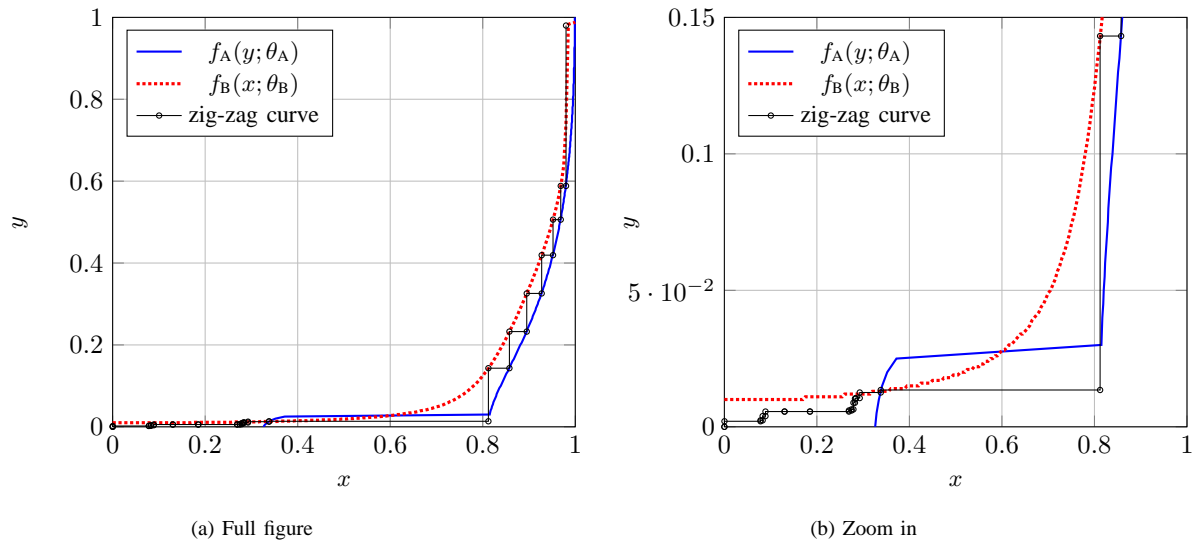

Due to the relaxation, however, the outputs of the optimization solver may not
all be feasible for the original optimization.  For example, in
\eqref{eq:conexample}, even when the $M$ interpolated points satisfy
the inequality, it is possible that there exist some other points in
the line segment $(x_{t-1},x_t]$ violating the inequality.  This tends
to happen especially when $x_t-x_{t-1}$ is large.
Fig.~\ref{fig:infeasible} illustrates such an example. The region
below the dotted curve and left of the solid curve is
$(\theta_\va,\theta_\vb)$-feasible. But there are two disjoint
$(\theta_\va,\theta_\vb)$-feasible regions. The first intersection of
these two curves is not the target point $(\eta_\va,
\eta_\vb)$. 
Fig.~\ref{fig:bp-decoding} plots the curves for a feasible output of
the optimization solver for the same values of $\alpha_\va$, $\alpha_\vb$,
$\eta_\va$ and $\eta_\vb$, where the degree distributions are
\begin{IEEEeqnarray*}{rCl}
\Psi_\va(x) & = & 0.1040 x +0.8362 x^2+0.0582 x^{26} + 0.0007
x^{27}, \\
\Psi_\vb(x) & = & 0.1133 x + 0.7902 x^2 + 0.0662 x^{13} + 0.0284 x^{14} +
0.0020 x^{15}.  
\end{IEEEeqnarray*}

Therefore, we need to verify the feasibility of each output of the
optimization solver. Though it is possible to increase the chance of
obtaining feasible outputs by using larger values of $M$ and
$t_{\max}$, the optimization solver will run longer time. For example,
we use $M=20$ and $t_{\max}=20$ for the results in Table~\ref{tab:1}.
Instead of using larger $M$, we add constraints to avoid a large jump
from $x_{t-1}$ to $x_t$.  For those outputs that are not feasible, we
may reduce the value of $\theta_\va$ and $\theta_\vb$ a little bit to
make the solution feasible.

Since those optimization problems are non-convex, we may not obtain
the global optimal value. Therefore, we run the optimization solver
multiple times (with randomly selected initial point) and pick the
best among all the outputs. For the parameters we have
evaluated, the feasible outputs returned by the optimization solver
are all very close to the theoretical upper bound.


\begin{thebibliography}{10}
\providecommand{\url}[1]{#1}
\csname url@samestyle\endcsname
\providecommand{\newblock}{\relax}
\providecommand{\bibinfo}[2]{#2}
\providecommand{\BIBentrySTDinterwordspacing}{\spaceskip=0pt\relax}
\providecommand{\BIBentryALTinterwordstretchfactor}{4}
\providecommand{\BIBentryALTinterwordspacing}{\spaceskip=\fontdimen2\font plus
\BIBentryALTinterwordstretchfactor\fontdimen3\font minus
  \fontdimen4\font\relax}
\providecommand{\BIBforeignlanguage}[2]{{%
\expandafter\ifx\csname l@#1\endcsname\relax
\typeout{** WARNING: IEEEtran.bst: No hyphenation pattern has been}%
\typeout{** loaded for the language `#1'. Using the pattern for}%
\typeout{** the default language instead.}%
\else
\language=\csname l@#1\endcsname
\fi
#2}}
\providecommand{\BIBdecl}{\relax}
\BIBdecl

\bibitem{Zhang06}
S.~Zhang, S.~C. Liew, and P.~P. Lam, ``Hot topic: Physical-layer network
  coding,'' in \emph{Proc. MobiCom '06}, New York, NY, USA, 2006.

\bibitem{Nazer11}
B.~Nazer and M.~Gastpar, ``Compute-and-forward: Harnessing interference through
  structured codes,'' \emph{Information Theory, IEEE Transactions on}, vol.~57,
  no.~10, pp. 6463--6486, 2011.

\bibitem{lu13n}
\BIBentryALTinterwordspacing
L.~Lu, L.~You, and S.~C. Liew, ``Network-coded multiple access,'' 2014,
  accepted by IEEE Trans. Mobile Computing, early access available at IEEE
  Xplore. [Online]. Available: \url{http://arxiv.org/abs/1307.1514}
\BIBentrySTDinterwordspacing

\bibitem{ncma14}
\BIBentryALTinterwordspacing
L.~You, S.~C. Liew, and L.~Lu, ``Network-coded multiple access {II}: Toward
  realtime operation with improved performance,'' 2014. [Online]. Available:
  \url{http://home.ie.cuhk.edu.hk/~yl013/docs/NCMA2draft.pdf}
\BIBentrySTDinterwordspacing

\bibitem{cocco2014}
G.~Cocco and S.~Pfletschinger, ``Seek and decode: Random multiple access with
  multiuser detection and physical-layer network coding,'' in
  \emph{Communications (ICC), 2014 IEEE International Conference on}.\hskip 1em
  plus 0.5em minus 0.4em\relax IEEE, 2014.

\bibitem{verdu1998multiuser}
S.~Verdu, \emph{Multiuser detection}.\hskip 1em plus 0.5em minus 0.4em\relax
  Cambridge university press, 1998.

\bibitem{feng2013algebraic}
C.~Feng, D.~Silva, and F.~Kschischang, ``An algebraic approach to
  physical-layer network coding,'' \emph{Information Theory, IEEE Transactions
  on}, vol.~59, no.~11, pp. 7576--7596, Nov 2013.

\bibitem{feng2013communication}
C.~Feng, R.~W. N{\'o}brega, F.~R. Kschischang, and D.~Silva, ``Communication
  over finite-ring matrix channels,'' in \emph{Information Theory Proceedings
  (ISIT), 2013 IEEE International Symposium on}.\hskip 1em plus 0.5em minus
  0.4em\relax IEEE, 2013, pp. 2890--2894.

\bibitem{lubyLT}
M.~Luby, ``{LT} codes,'' in \emph{Proc. 43rd Ann. IEEE Symp. on Foundations of
  Computer Science}, Nov. 2002, pp. 271--282.

\bibitem{shokRaptor}
A.~Shokrollahi, ``{R}aptor codes,'' \emph{Information Theory, IEEE Transactions
  on}, vol.~52, no.~6, pp. 2551--2567, Jun. 2006.

\bibitem{shengli09}
S.~Zhang and S.-C. Liew, ``Channel coding and decoding in a relay system
  operated with physical-layer network coding,'' \emph{Selected Areas in
  Communications, IEEE Journal on}, vol.~27, no.~5, pp. 788--796, June 2009.

\bibitem{wubben10}
D.~Wubben and Y.~Lang, ``Generalized sum-product algorithm for joint channel
  decoding and physical-layer network coding in two-way relay systems,'' in
  \emph{Global Telecommunications Conference (GLOBECOM 2010), 2010 IEEE}, Dec
  2010, pp. 1--5.

\bibitem{liew2013physical}
S.~C. Liew, S.~Zhang, and L.~Lu, ``Physical-layer network coding: Tutorial,
  survey, and beyond,'' \emph{(invited paper) Physical Communication}, vol.~6,
  pp. 4--42, 2013.

\bibitem{zhu2014isit}
J.~Zhu and M.~Gastpar, ``Gaussian (dirty) multiple access channels: A
  compute-and-forward perspective,'' in \emph{Information Theory (ISIT), 2014
  IEEE International Symposium on}, June 2014, pp. 2949--2953.

\bibitem{zhu2014multiple}
------, ``Multiple access via compute-and-forward,'' \emph{arXiv preprint
  arXiv:1407.8463}, 2014.

\bibitem{puducheri2007design}
S.~Puducheri, J.~Kliewer, and T.~E. Fuja, ``The design and performance of
  distributed {LT} codes,'' \emph{Information Theory, IEEE Transactions on},
  vol.~53, no.~10, pp. 3740--3754, 2007.

\bibitem{hern12}
B.~Hern and K.~Narayanan, ``Joint compute and forward for the two way relay
  channel with spatially coupled {LDPC} codes,'' in \emph{Global Communications
  Conference (GLOBECOM), 2012 IEEE}, Dec 2012.

\bibitem{casini2007contention}
E.~Casini, R.~De~Gaudenzi, and O.~R. Herrero, ``Contention resolution diversity
  slotted {ALOHA} ({CRDSA}): An enhanced random access schemefor satellite
  access packet networks,'' \emph{Wireless Communications, IEEE Transactions
  on}, vol.~6, no.~4, pp. 1408--1419, 2007.

\bibitem{liva2011graph}
G.~Liva, ``Graph-based analysis and optimization of contention resolution
  diversity slotted {ALOHA},'' \emph{Communications, IEEE Transactions on},
  vol.~59, no.~2, pp. 477--487, 2011.

\bibitem{paolini2011high}
E.~Paolini, G.~Liva, and M.~Chiani, ``High throughput random access via codes
  on graphs: Coded slotted {ALOHA},'' in \emph{Communications (ICC), 2011 IEEE
  International Conference on}.\hskip 1em plus 0.5em minus 0.4em\relax IEEE,
  2011, pp. 1--6.

\bibitem{stefanovic2012frameless}
C.~Stefanovic, P.~Popovski, and D.~Vukobratovic, ``Frameless {ALOHA} protocol
  for wireless networks,'' \emph{Communications Letters, IEEE}, vol.~16,
  no.~12, pp. 2087--2090, 2012.

\bibitem{narayanan2012iterative}
K.~R. Narayanan and H.~D. Pfister, ``Iterative collision resolution for slotted
  aloha: An optimal uncoordinated transmission policy,'' in \emph{Turbo Codes
  and Iterative Information Processing (ISTC), 2012 7th International Symposium
  on}.\hskip 1em plus 0.5em minus 0.4em\relax IEEE, 2012, pp. 136--139.

\bibitem{paolini2014coded}
E.~Paolini, G.~Liva, and M.~Chiani, ``Coded slotted {ALOHA}: A graph-based
  method for uncoordinated multiple access,'' \emph{arXiv preprint
  arXiv:1401.1626}, 2014.

\bibitem{cover06}
T.~M. Cover and J.~A. Thomas, \emph{Elements of Information Theory},
  2nd~ed.\hskip 1em plus 0.5em minus 0.4em\relax John Wiley \& Sons, Inc, 2006.

\bibitem{luby98}
M.~Luby, M.~Mitzenmacher, and M.~A. Shokrollahi, ``Analysis of {R}andom
  {P}rocesses via {A}nd-{O}r {T}ree {E}valuation,'' in \emph{SODA}, 1998, pp.
  364--373.

\bibitem{Raptormono}
A.~Shokrollahi and M.~Luby, \emph{Raptor Codes}, ser. Foundations and Trends in
  Communications and Information Theory.\hskip 1em plus 0.5em minus 0.4em\relax
  now, 2011, vol.~6.

\end{thebibliography}
\end{document}